# Cluster of Gamma-Ray Bursts – Image of a Source. Catalog of Clusters (Sources) of Gamma-Ray Bursts.

A.V. Kuznetsov

Space Research Institute, 117810, Profsouznaya 84/32, GSP-7, Moscow, Russia

**Abstract**. The clusters of gamma-ray bursts are considered which are assumed to be images of the repeated gamma-ray burst (GRB) sources. It is shown, that localization of the cosmic gamma-ray burst sources (GBS) is determined by the clusters of GRBs. About 100 candidates in sources are presented in the form of the catalog, which is compiled relying on the base of the BATSE data up to middle of 2000. Gamma-ray bursts (from 5 to 13) of a cluster that display a source do not coincide in their position. The catalog table containing basic information about the GRB sources yields the possibility to research the GBS properties and their identification. The birth of GRBs in the clusters allows predicting the appearance of GRBs both in time and space. Most general properties of the supposed GRB sources are discussed. An attempt to compile the first GRB source catalog is made.

## 1. Introduction.

Before BATSE experiment the possible time of repetition of classical gamma-ray bursts was estimated based on some model concepts in [1, 2]. Which imply that the upper limit may be 10 years, whereas the lower limit may be several months. During the BATSE era it became possible to solve the problem of a source repeating manifestation. The papers have been appeared which consider the existence of GRB which are clustered in space and time. It is supposed, that these events represent one source. Quashnok and Lamb [3] at study angular distribution of the nearest by the position GRBs of the 1B catalog find repeated events on space ~ 5 angular degrees and assume existence of classical bursts with time of recurrence about several months. Wang and Lingenfelter work [4] was one of the first, where five classical gamma-ray bursts also from the catalog 1B were considered, as the repeated events, radiated from a source, for which were determined coordinates. Brainerd et al. [5] investigated the 2B catalog, using time-dependent 2-point correlation function, and they have not found the confirmation of recurrent bursts. In result of more detailed study of the 2B catalog Meegan et al. [6] also come to a negative conclusion. They determine the top limit on repeated classical GRBs, which makes no more than 20% from total number on temporary scale about one year. Then Tegmark et al. [7] investigate possible clusters on any angular scale and time interval ≥ 1 year on the data of the 3B catalog, containing 1122 GRBs, using the spherical harmonic analysis. Their conclusion: no more than 5 % of bursts BATSE can be connected with sources of repeated events. Currently the concept of the lack of repeating GRBs is in fact universally accepted.

Here it should be determined that those GRBs are assumed to be repeating that have identical coordinates. In this case the dimension of GRB cluster does not exceed the error box of individual gamma-ray burst. Generally, as in above-mentioned papers, it is assumed that the coordinates of a GRB and its source coincide and a search for repeating events actually means that of repeating manifestation of source.

At the same time paper [8] shows the possibility of existence of GRB clusters within the fixed area of finite dimension, which exceeds the error box value of burst. The gamma-ray bursts in these clusters do not coincide in position, and in this sense they are not repeating bursts. Indeed, the cluster dimension points to the fact that the positions of bursts forming a cluster cannot coincide. The gamma-ray bursts of above discussed clusters are singled out by sign of their belonging to the same space-time region.

For the BATSE in the area with r = 12 angular degrees corresponding to the solid angle 0.01 of 4π in the average about three GRBs per year or one GRB for four months are observed. However, up to 10 events for the same 4 months are sometimes detected. The reality of these clusters leads to the evident though unexpected conclusion – localization of a single event does not determine coordinates of the source. Its position (and its properties) is determined by some great number of GRBs. The above discussed clusters result in paradox associated on one hand with the lack of repeating events, which coincide in position (that is in good agreement with above results of BATSE data researches), on the other hand the cluster of GRBs are repeating in the sense that they belong to the same source.

The well-known cluster of 4 events (GRB 961027.1: α,β = 67°.5 - 42°.4; GRB 961027.2: α,β = 68°75 - 54°.3; GRB 961029.1: α,β = 59°.5 - 52°.6; GRB 961029.2: α,β = 59°.75 - 48°.9) observed from the same region during 1.8 days [9] may be taken as an example. The domain of cluster is determined by a circle 6.1° in radius with the center at point l”, b” = 255°.2 - 43°.9 (see Table 1). Statistical estimation [10] with high reliability points that manifestation of these 4 events correlated in time and space is not random. Interpretation of such events as sources leads however to obvious contradiction associated with the fact that GRBs clusters do not coincide in position with the point source. In the specific cluster of 4 GRBs when the cluster dimension and localization error are close in value these GRBs can be coincided with one point, for example, in the case of possible underestimating localization error. However, in general case when the characteristic dimension of a cluster exceeds the localization error region by a factor of three and more, the necessity to explain the observed cluster extent becomes evident.

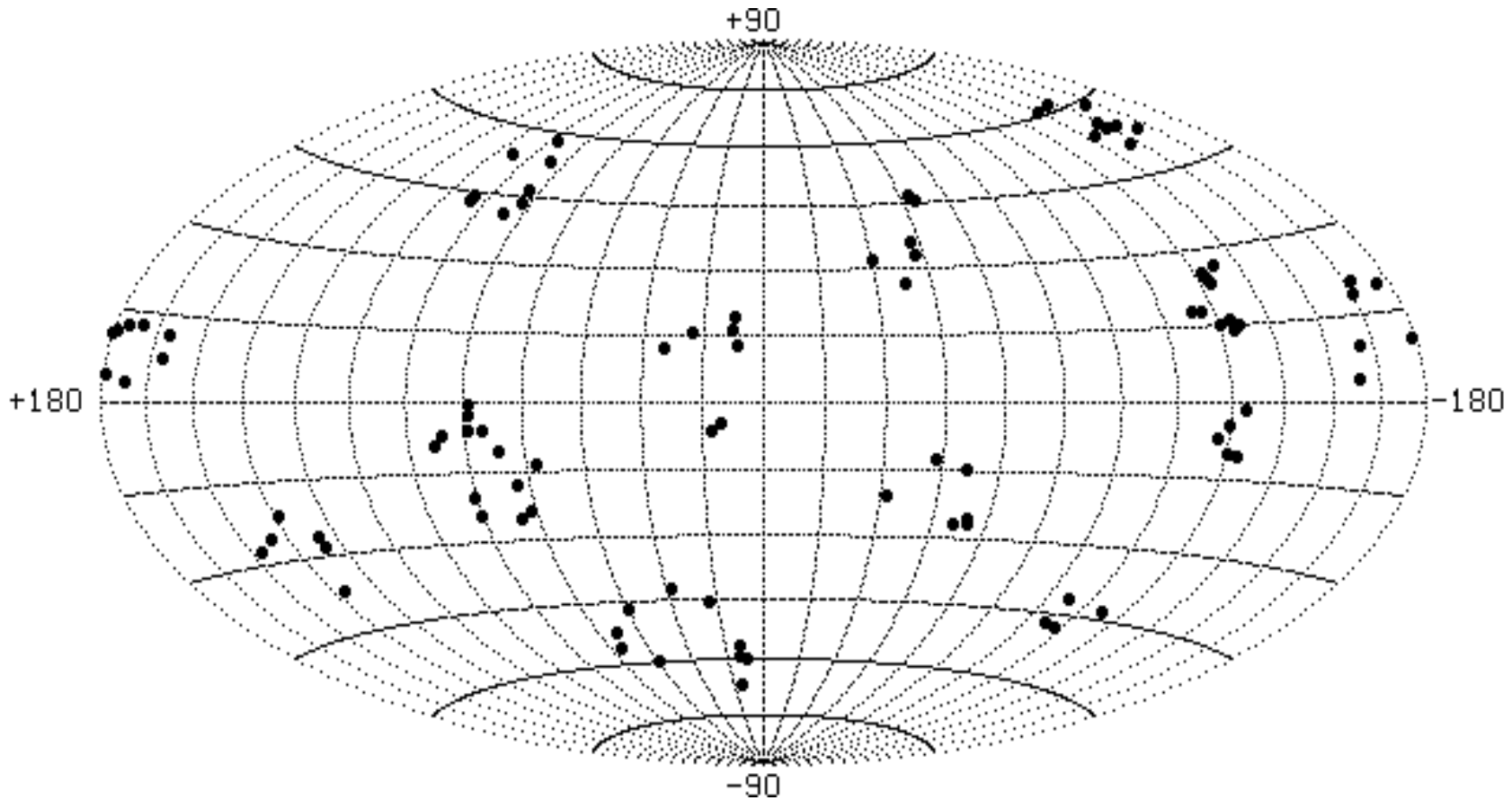

**Fig. 1**. The clusters of gamma-ray bursts.

Fig. 1 shows in galactic coordinates the several clusters with different angular dimensions and the number of bursts, including the mentioned cluster of 4 GRBs. The black circle in Fig. 1 shows as an illustration the GRB position with an error box of 2 degrees, whereas a real localization error for the BATSE in the average is about 4 degrees. Note that GRBs are not concentrated towards the center or some other cluster point, rather they are located near the boundary, often forming ring-like cluster.

The possibility to consider the observed clusters as GRB sources leads to the problem associated with a necessity of refusal from a usual point image of repeated GRB source.

This paper is the result of systematical search for GRB sources beginning from 1995 [11]. When compiling a preliminary catalog [8] it was shown that GRBs form clusters, which have the finite dimensions. Using the present data the angular sizes of GRB clusters are within 10-28 degrees, and the number of events in them ranges from 5 to 13. The lifetime (manifestation time) of clusters varies from several weeks to several months. Direct interpretation of the data led to the conclusion that the burst cluster with the large angular and appropriate linear sizes should be in the nearest space. Based on this, in general, logical assumption an attempt was made to consider heliospherical origin of GRB [8]. This attempt, however had failed.

Recent progress in studying the phenomenon is associated with the BeppoSAX experiment [12], which provided exact localization and detection of accompanying electromagnetic radiation within all range from X-ray to radio. In the context of measuring a considerable redshift in spectra of some events cosmological origin of GRBs is now the most preferable. But as long as a mechanism and an energy source of this extraordinary phenomenon are unknown we cannot eliminate Galaxy as space of GRB origin.

The paper involve the following sections:
- cluster selection procedure;
- statistical threshold for cluster selection;
- table of GRB sources;
- general properties of sources;
- discussion and conclusions.

Though there is no difference between the concepts cluster (source image) and GRB source in the text it is assumed that the cluster center and GRBs, mentioned in Table specify the GRB source position and its properties, respectively.

## 2. Cluster Selection Procedure.

Among 2820 GRBs detected for 9 years ranging from 21.04.1991 to 25.05.2000 we analyzed 2555 events with a localization error not higher than 10 angular degrees. The analysis was made using the BATSE 4Br catalog [13], and yet with regard to the importance of exact localization we use the BeppoSAX [14] on 1.01.01 and RXTE [15] data. The below-described selection makes it possible to identify the concrete GRB clusters. This methodology takes into consideration observation factors:

a) visible extent of a source, as well as concentration of burst in most clusters towards the external boundary necessitates a search for clusters using the coordinate grid rather than the bursts. The coordinate grid is formed with overlapping of cells over the longitude and latitude, that enables identification of the expected clusters already at the initial stage. (Actually all present programs are designed to detect GRB point sources);

b) GRBs of a cluster are radiated by the source during the relatively short time. The ratio of "active phase" duration to the time when the source is not seen is less than 1:10;

c) cluster boundary is described by the circle.

GRB clusters are singled out at two stages. At the first stage coordinate selection is made – GRBs are fixed in each cell of the grid. Selection is made over the latitude belts in the areas with r = 18 angular degrees with overlapping of the neighboring areas over the latitude and longitude at 18 degrees. Thus obtained Table contains the initial data and provides preliminary identification of concrete clusters and reliability of selection.

At the final stage each cluster, identified at the first stage is analyzed. Selection of actual clusters becomes possible due to the fact that they have natural spatial and time boundaries. At the areas the possible clusters of 5 and more bursts are selected, for which the radius and the center position are determined. The estimation of the cluster spatial boundary is given on Fig. 2. The empty ring around a cluster has an area of the same order that a cluster itself (actually ≥ 0.8 of its value) and in the most cases exceeds it.

The selection completes possible identification of the cluster core, minimization of cluster size. The necessity of this operation is dictated by the fact that sometimes GRBs are located compactly in time or (and) space, and for this reason have low random probability. Than it is possible to illegally extend the cluster size due to inclusion of GRBs, which are background relative to the given cluster.

As a result for each supposed cluster its lifetime, center coordinates and its minimal size are determined.

The conclusion about existence of cluster is accepted according to the estimation of probability of its random formation based on binomial distribution or Poisson statistics.

The Fig. 2 illustrates the characteristic picture of anisotropy of angular GRBs distribution (if the cluster exists), where the origin of coordinates coincides with the cluster center (l'', b'' = 99.°8 – 56°.0). The dotted line shows the calculated number of GRBs as a function of a spatial angle in the isotropic case for cluster lifetime T.

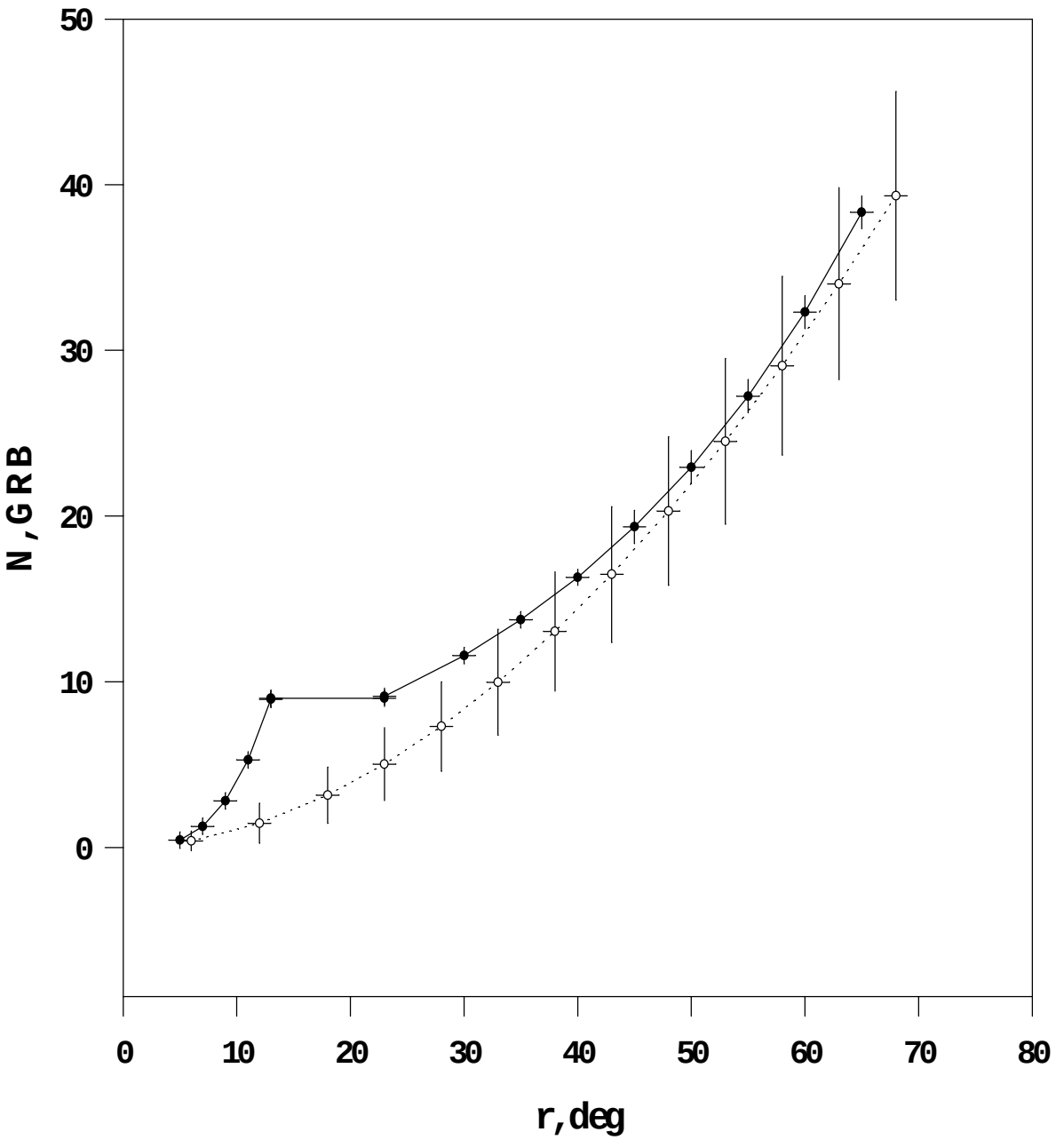

**Fig. 2**. GRB angular distribution determined by a cluster.

A total number of GRBs on the sphere corresponds to observed number N of GRBs during the same fixed time.

The solid line is a real distribution under existence of the cluster consists of three parts:

- the first one corresponds a GRB cluster and is characterized by a sharp increasing;
- the horizontal part which do not comprise any burst and is a empty band allowing to easily determine the cluster boundary;
- the third part corresponds to the calculated (isotropic) curve to which it gradually approaches and finally coincides.

The presented cluster involves 9 events. The circumference with 13.1°, comprising events, occupies 0.013 of the entire sphere, whereas the boundary band without events takes only 0.011 of ones.

Under identification of cluster in 100% "vacuum" around a clusters is essentially observed. Another words there is the absence of background clusters over angular space equal to the cluster area, where 1 or 2 events are expected. In such a case we should assume that the absence of background is not random and consider GRBs of cluster as a pure effect. The absence also assumes that all GRBs are generated in clusters.

The presented catalog comprises 618 GRBs (25% of a total number of the analyzed bursts) united in 94 clusters. It is well to bear in mind that the BATESE instrument is able to observe only 1/3 of a celestial sphere. This sky coverage coefficient $k_s$ is determined by persistent 35 % celestial sphere shadowing by Earth, by deadtime associated with South Atlantic anomaly passage, by data recording time into instrument memory and by other reasons. Taking into account these factors limited BATSE "sensitivity" we must assume that the 1st catalog involving 260 GRBs, can really consist of around 800 GRBs [16].

Incomplete and variable coverage of the celestial sphere would distort the source analysis results. This distortion results in the decreasing of observed GRBs from the source and hence the decreasing the number of detected sources.

## 3. Statistical Threshold for Cluster Selection.

Identification of clusters, which reveal small-scale angular anisotropy in GRB distribution inevitably, brings about the question about their reality that is generally confirmated by statistical estimation. This estimation is carried out in terms of observed large-scale isotropy of GRBs. For some clusters as for above discussed GBS of 4 GRBs, which are irradiated from the same region during 1.8 days, it is obviously that the probability of their random formation is low [9]. It concerns properly any real cluster, a typical example of which is given in Fig. 2.

It is essential that the source observations (BATSE experiment) are made only during some part $\tau$ of an active phase time $\Delta T$ determined by the sky coverage coefficient ($k_s$): $\tau = k_s \cdot \Delta T$. The observation time dependence of a number of detected GRBs permits to estimate the real number of bursts in a cluster $n = N_r \cdot k_s$ where n – observed number of bursts, $N_r$ – their real number. The active phase is about several months; total observation time (T) is up to 9 years.

This selection technique uses alternating fixing of the space domain and cluster manifestation time for determination of their minimal values. The idea of statistical estimation is seen from Fig. 2, where the manifestation of cluster is shown for the fixed time. Identification of GRBs clusters is based on determination of minimal radius of circumference (solid angle) contained cluster bursts, and minimal time during which these events are observed. This method eliminates the estimation problem *a posteriori* [10] since it prevents under analysis violation in the choice of the parameter values analysis, which are set by the initial data themselves.

As it is shown in [17] two main types of models of repeating GRBs sources should be assumed in analyzing the observation data: a) stochastical model, when the source randomly radiates GRB with a fixed velocity, so the probability of burst observation within any given time interval is persistent. The number of bursts from the i-st source would be distributed according to Poisson; b) the source episodically radiates GRBs, for example during the active phase, with the time $\Delta T$. In this case the number of observed bursts from the source obeys binomial distribution.

The considered cluster - GRB sources are obviously related to second model.

Statistical criterion of cluster identification is based on binomial distribution which implies the probability $p_s$ of detecting m GRBs at the given solid angle $p_s = C^m{}_n \, p^m \, q^{n-m}$. The calculated probability taken as a statistical threshold for cluster selection is $\leq 3 \cdot 10^{-3}$, that corresponds to $3\sigma$ in Gauss statistics. The feasibility of using this estimation approach for Poisson statistics in identification of GRB cluster is shown in paper by Wang and Lingenfelter [4].

Since the probability of detecting a cluster increases as the observation time increases, the $p_s$ should be multiplied by the number of time intervals corresponding to the cluster lifetime, which meet observation time of given region. The final value of random probability $p = p_s \cdot k_t$ , where $k_t$ is the number of time intervals during the time before repeating observation of the source. According to the data (see Table 1) the time between repeating manifestation of the source ranges from 0.5 to 5 years and more, i.e. in the average it is 3-4 years, and the mean lifetime of a cluster is 3-4 months. Following to this $k_t$ =10 is taken for calculations. The cluster of GRBs is considered to be real if the probability value of its random formation calculated from the binomial distribution formula $p_s \leq 3 \cdot 10^{-4}$. Table 1 gives the calculated probability values in terms of $k_t$ coefficient.

For instance, using the actual data and the expression for expected random number of bursts $\mu = m(1- \cos r)/ 2$ we expect to have 1 burst for 4 months at the area with $r = 12°$ under observation m = 100 bursts on the entire sphere for the same 4 months. To obtain the random probability $p_s$ less then $3 \cdot 10^{-4}$ the additional 6 events should be observed in this region during the same time.

Note that the mentioned threshold may be highly exceeded since it disregards the sky coverage coefficient $k_s$, which for the BATSE is about 1/3 and specifies the probability that GRB will be missed.

## 4. Table of GRB Sources.

Table 1 gives a catalog of supposed GRB sources detected from 21.04.91 to 25.05.2000, which is far from complete. The estimation shows that for that time at the BATSE sensitivity level 7000 events must be detected for total sky sphere. A loss of number of events in clusters leading to the loss of clusters is primarily associated with the limited field of view, variability of the observation region and interruption in instrument operation. The mentioned reasons bring about the fact that among the total number of 2555 GRBs only 618 can be detected in clusters. It must be kept in mind that in majority of real cases the loss of one GRB result in the loss of whole cluster.

Table 1 shows clusters of 5 and more GRBs, which are defined the main properties of 94 GRBs sources. Column 2 gives the time of observing gamma-ray bursts from this source, which reflects the duration of a source active phase. Column 3 shows consequently the source center coordinates for two systems: galactic and equatorial. Two latter parameters are basic to designate a source. Obviously the most accurate designation of a source would be described simultaneously by coordinates and time of its observation. For example, the source from Table 1 with ordinal number 1 should be designated as GBS: l, b = 5.6-10.2 (981031-990216) for galactic coordinate system. In column 4 a digit after a point in GRB designation shows its ordinal number for the given day. Columns 5 and 6 give coordinates of GRBs and their localization accuracy. Column 7 shows the angular distance of given burst from the source - cluster center. Here a maximal radius of cluster specifies its angular size. The latter column gives also the values of probability of random formation of a cluster (p) and significance level n$\sigma$ at which the given cluster is selected.

## 5. Common Properties of GRB Sources.

Having information about almost one hundred GRB sources we can make preliminary conclusions in respect of their most common properties. The catalog Table presents, at our opinion, the most evident GRB clusters. Beyond any doubt a majority of them are real and obtained results may be practically used.

The sources have characteristic parameters among which the angular cluster size (10-28 degrees) is more important. Today hardly it is possible to consider seriously the existence in the Universe of astronomical objects with angular sizes of about 20°. In this connection it should be assumed that the angular size of a cluster does not define the source real size. And we are to speak only about the image of GRB source, which can be created, for example, by local space-time curvature.

The cluster boundaries are described by the circumference, observed symmetry points to real source position in the cluster center.

The 94 selected sources comprise 618 events or about 25% of all analyzed GRBs. Consideration of all factors decreasing a number of detected clusters which involve variable field of view, dead time, telemetry malfunctioning, edge effect, exclusion from analysis 4 GRB clusters and so on, would result in increasing a number of events in clusters. Rough estimation, which takes into account only clusters of 4 GRB, field of view variability and edge effect leads to increasing of bursts in clusters to 65%. This and some other estimations and considerations allow to assume that at least a majority of GRBs form clusters.

The background lack connected with empty space around a cluster yields independent conclusion that bursts are generated in clusters.

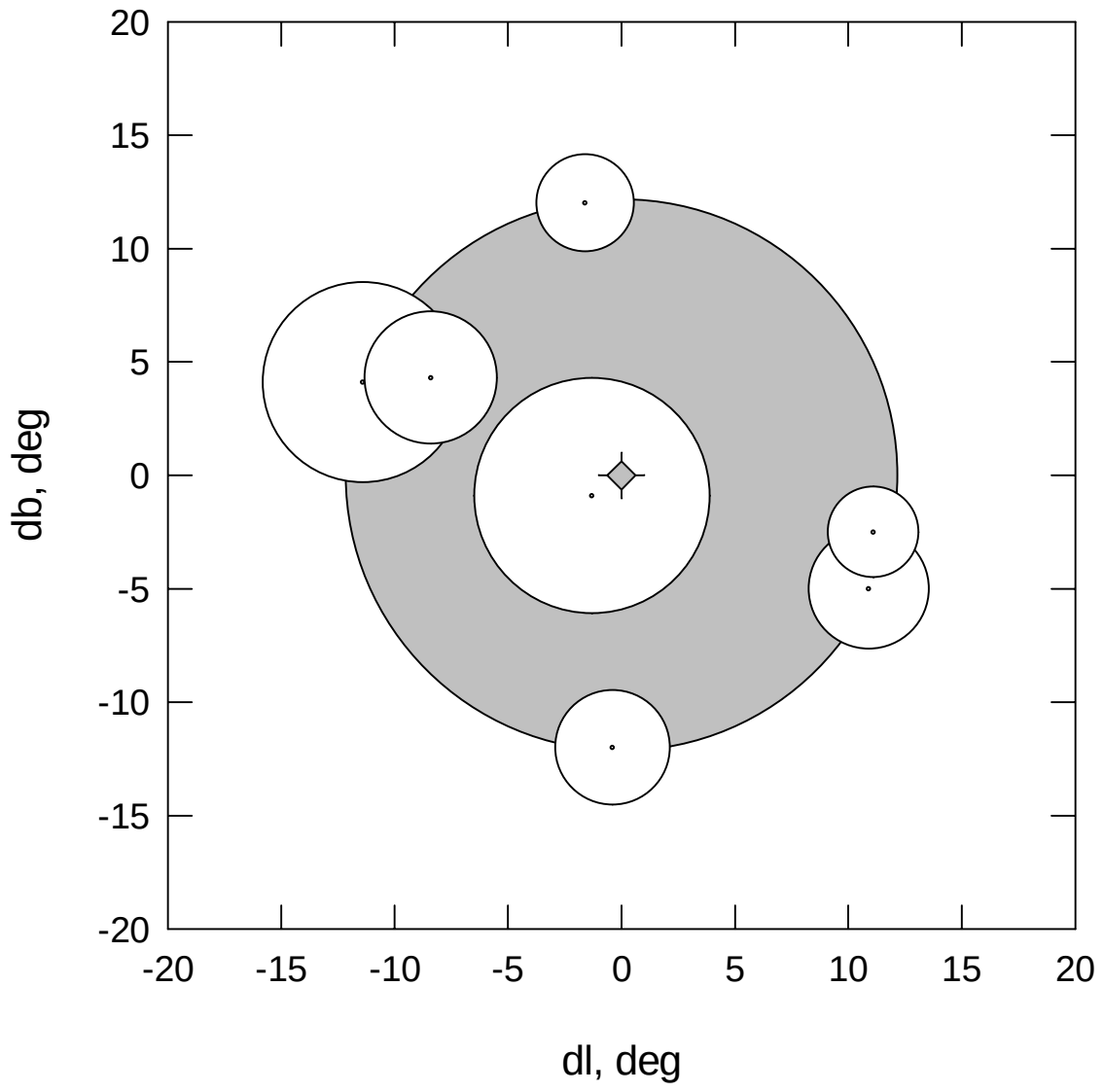

**Fig. 3**. The typical picture of GRB distribution in a cluster.

In the context of assumption that a cluster has a ring-like shape possible deficit of GRBs near the center was estimated based of uniform distribution of events over the solid angle. All clusters were normalized to one size. In the region with 5-arc degrees radius, where 79 GRBs out of a total number of 618 are expected, in reality only 37 are located. The deficit is observed at the significance level ~ 5σ. This estimation indicates that GRBs mainly avoid the cluster center, but a large error box does not allow to determine events, which are maximal close to the center.

One of the most practically important conclusions of this paper is the possibility to predict on basis cluster properties GRB appearance in time and space. One current data provide prediction the observation time of events in the given cluster that involves not less than 5 events, which fairly well determine the center position and size.

In the specific case the prediction of GRB appearance in space is restricted by the cluster size, whereas in time it is restricted by the interval between the events, which averages about two weeks. The use of prediction is essential for observation of afterglow from the GRB source.

Fig. 3 gives the cluster of 7 GRBs with the center coordinates l, b = 314°.0 – 3°.9 (№ 78) and helps to understand how the source center position, cluster size and their errors are determined. Shadowing shows the cluster size that is the circumference with a minimum radius (in the given case r = 12°.1), which is determined by the boundary GRB position. The asterisk shows the cluster center. Remind that cluster center (source position) is calculated with respect to the minimum radius circumference involved all GRBs of the given cluster. The error boxes of boundary events determine the error box of a source position. For this reason the error box of cluster center depends on a maximal localization error of external GRBs. For the cluster under consideration the accuracy of center position is 4°.2. With such an approach the source coordinates in the average are determined with an accuracy of about 6° as the initial data - GRBs have a maximal error box 10° and their average error is about 4°.

Assuming that clusters represent images of repeating GRB sources one can attempt to change from an image to a real source that is to restore the initial position of the source from GRBs localization. In simplest case an image is obtained from overestimation of the localization error. Then, for example, for a clustering on Fig. 3 an increase in the error maximum by 10° would bring about the coincidence of source position with the cluster center. However we should refuse from this solution for a number of reasons, one of which is the lack of cluster GRB concentration toward the center in all observed formations. We should probably proceed from the concept that the obtained source images are real. Decreasing the angular distance of each GRB from the center by the same factor (in the given case – 5.8) the position of all gamma-ray bursts would coincide with one point – a cluster center. Due to the symmetry of GRB distribution in a cluster the source position will evidently coincide with the cluster center for any model.

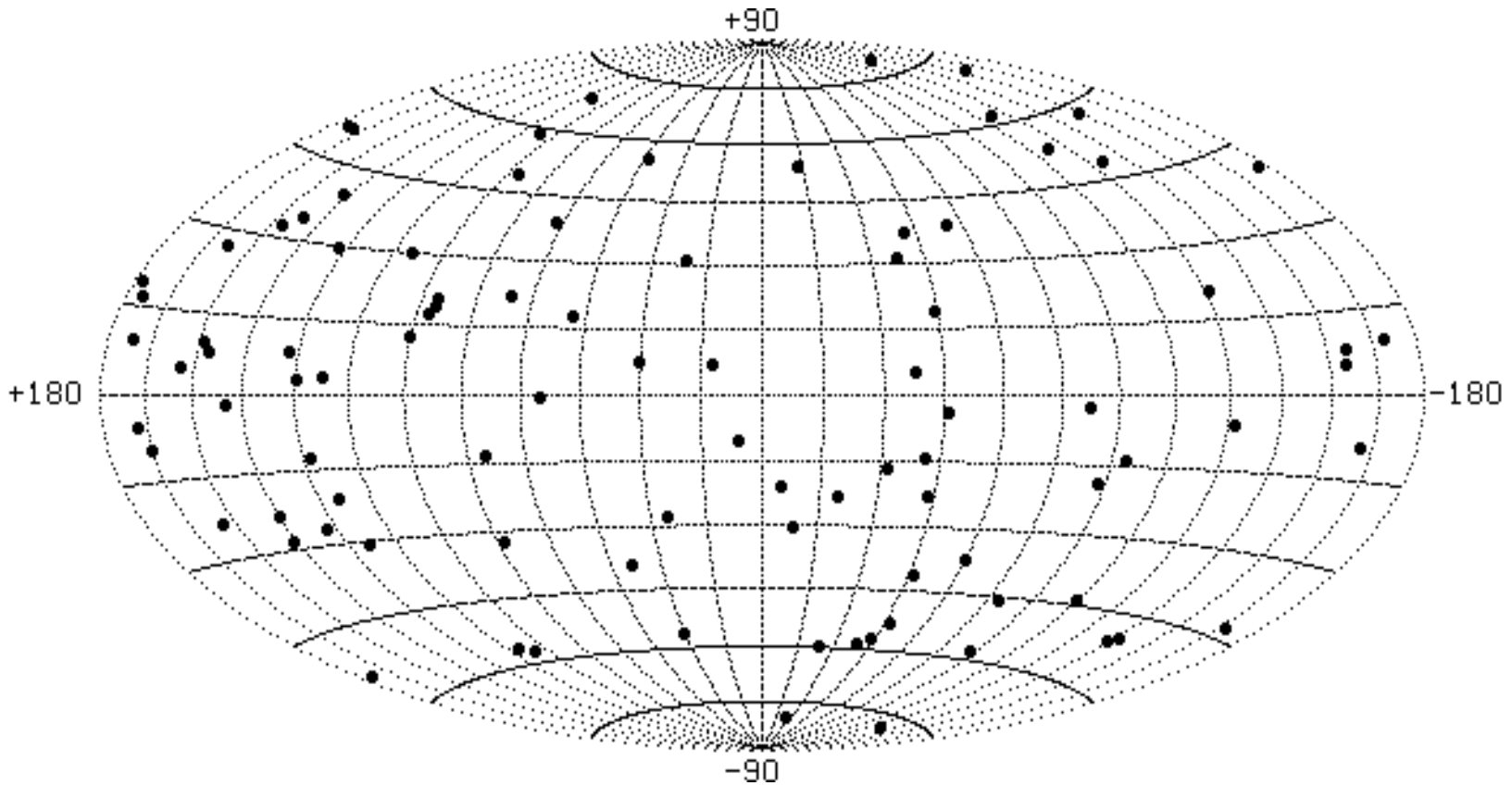

**Fig. 4**. GRB clusters on the sky sphere.

The existence of clusters which represent the image of a source in principle exclude observation of repeating events, because they do not yield information about the source position even though repetition is realized with high precision. The term itself "repeating gamma-ray burst" loses its meaning. At the same time we see that the majority of all repeating events (within the error region) is in clusters, and they are about a half of events of each cluster. For example Fig. 3 shows a cluster of 7 GRBs 4 of which form two repeating pairs. These events could hardly be explained by only random superposition. There is possibly another reason. Simple counting of repeating events in clusters, when the error box of one event covers the center of another event, results in more than 240 gamma-ray bursts. This value is about 40% of all events in clusters and probably is responsible for the upper limit to repeating GRBs about 20 %, found in [6].

Positions of 94 GRB sources over the sky sphere are summarized in Figure 4. The character of distribution does not contradict to the known isotropy of gamma-ray bursts that evidently should be expected. The main feature of this distribution is associated with repeating manifestation of GRB sources. The 5° value was taken as a maximal distance between the centers of GRBs sources to identify those, which are repeatedly manifested. Among 94 sources 24 are repeating ones that is more than 25% of their total number, and 1 source is observed 3 times. The time between repeating manifestation source of GRB detected for 9 years of observations ranges from 0.5 to 5 years. The mean value of time of GRB source repeating manifestation may be roughly estimated as 4 years.

For 94 sources the probability p(m) of random manifestation of repeating source (m=2) within the region with angular size 5°, calculated from binomial distribution, is equal to $2\cdot10^{-3}$. This estimation compels to assume that the observed repetition of GRB sources do take place.

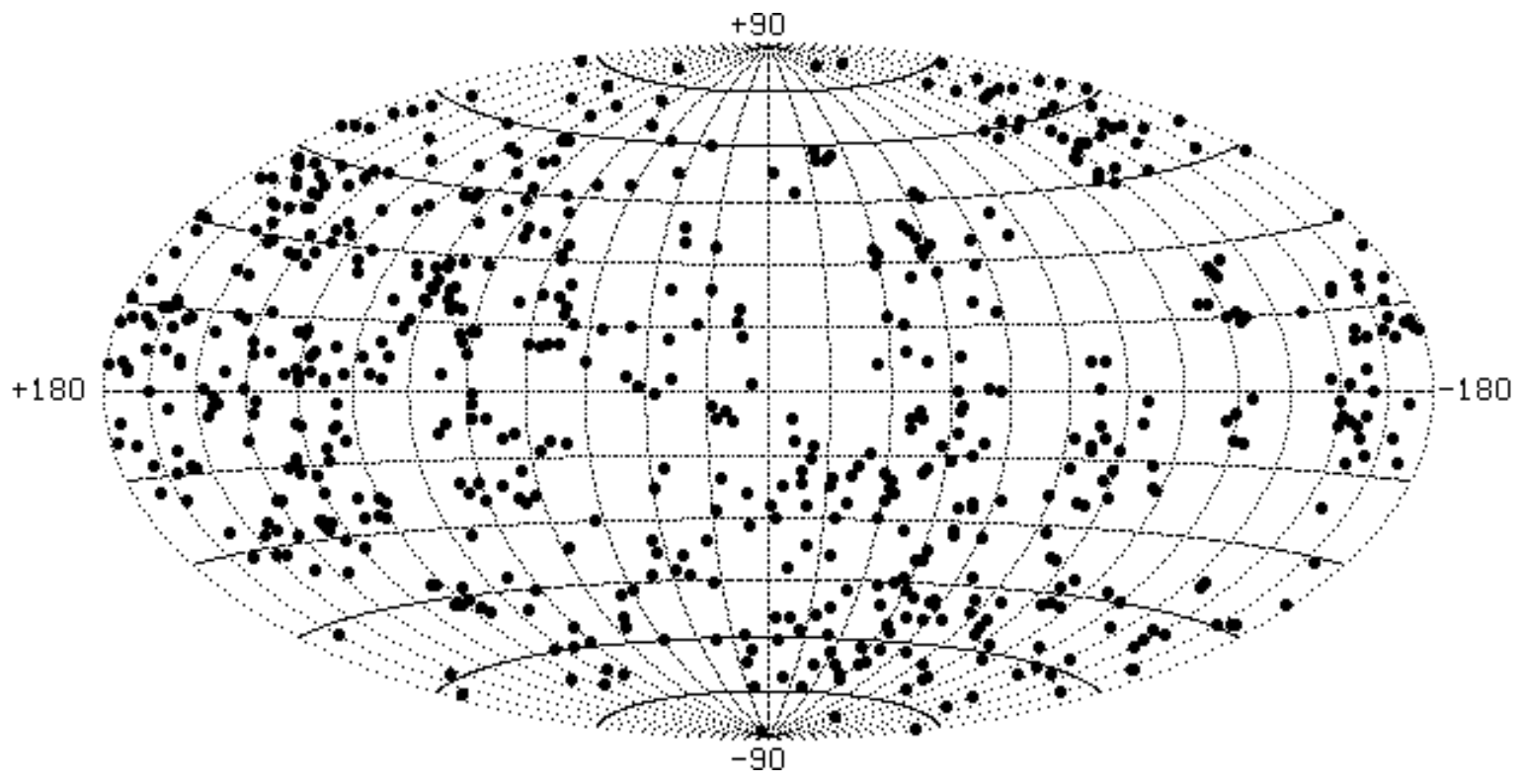

**Fig. 5**. 618 GRBs from 94 sources.

Fig. 5 gives the galactic coordinates distribution of 618 GRBs shown in Table 1 and united in 94 sources. There is a deviation from isotropy (white spots, which are turned out to locate in the wide band along the Earth equator). The deviation is induced by the known deficit of GRBs that have been recorded by the BATSE experiment in this band, due to shadowing by the Earth [13]. GRB clusters in this case are virtually used as an instrument of detection anisotropy of GRB source distribution.

The sources with practically the same localization error, which is about 6 angular degrees, should be used to identify them with the known astronomical object. The results suppose that we fix real a position of the source determined by GRB clusters as distinct from the existing concept "one GRB – one source". In terms of this conclusion we may estimate that the discrepancy between the position of GRB and that of the real source of this GRB may reach about 14 angular degrees.

The possibility of using the catalog to identify GRB sources is shown as an example of such interesting for the GRB problem astronomical objects as nearest galaxies, candidates for black holes and supermassive black holes. Thus, the nearest galaxies M31, M33 and M51 may be considered as potential GRB sources. M31 is identified with the source № 27 (l, b = 114°4 - 21°.3), M33 – with the source № 34 (l, b = 135°.8 - 28°.3), M51 – the source № 24 (l, b = 100°.9 + 68°.3). Discrepancy in coordinates ranges from 1.4 to 6 °. At the same time an attempt of identification of GRBs source with 5 candidates for supermassive black holes (M31, M87, M106, NGC 4261, NGC 4258) did not yield positive results, of course if M31 is excluded. Among 24 candidates for black holes in low - mass binary systems [18] GRO J0422 + 32 (N518 Per) can be singled out, that is identified with repeating source №№ 44,45: l, b = 165°.4 - 9°.6; l, b = 167°.9 - 5°.7 and GS 1354-645 (CenX-2) – № 78: l, b = 314°.0 - 3°.9.

## 6. Discussion and conclusions.

Detection of cosmic GRB clusters and their common properties bring about the results, which can form the basis of the comprehensive research of GRB source characteristics.

The observations suggest that the characteristic feature of gamma-ray bursts is their generation in the clusters, which are responsible for the localization and properties of their source. Estimation of a total number of bursts in clusters and the lack of background events confirm this conclusion.

GRB generation in clusters puts the source localization correctness in doubt. An evident condition is currently taken into account that localization of GRB within the error box coincides with that of source that is commonly fulfilled in observation a signal from the star object. The paper shows that such a concept may be wrong, and localization of a single GRB even with high accuracy would not be of practical value in determining location of its source. In this case independently of accuracy of GRB localization the error of a source coordinate determination may rich 14 degrees.

Correct localization is determined by a cluster of a 5 and more GRBs, that can be detected when observing the source region with an angular dimension of ~ 30 degrees continuously during several months.

The cluster lifetime corresponds to model of a source emitting bursts in the active phase. The 9 year observations established that active phase duration averages several months, the expected number of events for that period is close to maximum – about 20, whereas repeating manifestation of the source ranges from 1 to 5 years and more. Observation of repeating sources may be considered as an argument in favor of Galactic origin.

Observation of the entire celestial sphere (or its part) is permanently one of the basic conditions of GRB source images detection. For experiments it would be essential to find a compromise solution that allows us decreasing an accuracy of burst localization to increase at this expense of GRB detection efficiency – sensitivity and field of view of an instrument.

Generation of GRBs in clusters and properties of the latter makes it possible to predict GRB appearance in time and space. For example, according to the BATSE data it is possible a prognostication the time of event appearance from the given source on the base of most general properties of clusters. Detection of predicted event enables one to improve cluster characteristics and observe in the real time the GRB source region with the spacecraft and ground based instruments. The advantage of this observation is that ~ 20 GRBs may be really observed. This possibility is most essential for observations of GRB source afterglow, since, for example, in optical and radio bands the ground-based telescopes provide them.

Note that Table 1 with the cluster data is essentially an attempt to compile the first catalog of GRB sources.

A concept of GRB source image can be crucial in the problem of cosmic GRB origin. Practical value of GRB cluster properties is evident already now: GRB source identification conditions, possibility of prediction of GRB appearance in time and direction, provision of afterglow observation program from the Earth, requirements to the experiments on GRB observation and so on. At last, of practical results far from being complete, is realized irrespective of physical interpretation of the source image.

A search on the basis of results of this paper brings about possible identification of GRB sources with the nearest galaxies M31, M33, M51, M81 as well as with the representatives of a class of X-ray novae (BHXN) GRO J0422+32 and GS 1354-645.

The properties of GRB clusters were established from the observation data, and hence, should be compatible with any other experimental results. The most interesting is the compatibility of observed GRB clusters with the experimental data, which suggest that GRBs (some part) have cosmological origin. Recall, that the first and evident conclusion from cluster existence is that GRB and its source do not coincide by the position. This conclusion is at variance with GRB identification with “ host” galaxies. In this connection the validity of such identification puts in doubt. Determination of redshift from the optical afterglow spectrum has no obvious contradiction, but arises a question whether the interpretation of observed redshift as Doppler effect is correct. The problem of consistency becomes more interesting if it taken into account that GRBs with exact localization, afterglow and measured redshift are observed in clusters. For example, the well-known GRB 970508 for which the redshift $z = 0.835$ [19] was first measured, is an ordinary member of a cluster of a 7 events (№ 37 in Table 1). A total of 12 GRBs is observed in clusters, their exact localization was received with the BeppoSAX or with an interplanetary spacecraft network (IPN)[20].

Note that in case of feasible identification or any other observation search for GRB source we should not put in doubt its infinitely low (point) dimension. Speaking about an image of GRB source we have to assume that it is a result of unusual gravitational lensing.

The paper does not deal with a physical aspect of GRB origin, associated with the results obtained here, since it is of particular interest and will be discussed in the other paper.

The author is grateful to S.V. Repin for useful discussion and assistance.

**Table 1.** Catalog of the GRB sources (21.04.91- 25.05.2000).

| N | GBS life-time | GBS center, deg: l”, b” ($\alpha$,$\delta$) | GRB | l”, b”, deg | err, deg | $\theta$, deg | p nσ |
|---|---|---|---|---|---|---|---|
| 1. | 981031- | 5.6-10.2 | 981031 | 8.9- 7.1 | 2.0 | 4.5 | $3\cdot10^{-4}$ |
| | 990216 | (279.7-28.9) | 981203.2 | 12.0- 20.6 | 1.7 | 12.1 | 3.0 |
| | | | 981226.2 | 4.2+ 1.8 | 2.4 | 12.1 | |
| | | | 990112 | 14.2- 3.4 | 2.3 | 10.9 | |
| | | | 990128.1 | 354.3- 6.5 | 4.5 | 11.7 | |
| | | | 990216 | 353.4- 11.6 | 1.7 | 12.1 | |
| 2. | 920615- | 12.5+ 6.9 | 920615.1 | 6.5+13.0 | 6.6 | 8.5 | $3\cdot10^{-4}$ |
| | 921022 | (266.9-14.8) | 920707 | 10.5- 4.7 | 5.5 | 11.8 | 3.0 |
| | | | 920723.2 | 24.6+12.1 | 4.6 | 13.0 | |
| | | | 920808 | 17.9+15.5 | 3.1 | 10.1 | |
| | | | 920902.1 | 12.8- 6.2 | 2.8 | 13.1 | |
| | | | 920912 | 7.6+16.3 | 2.3 | 10.6 | |
| | | | 921022.2 | 7.4+18.9 | 1.7 | 13.0 | |
| 3. | 990206- | 20.8+30.8 | 990206.1 | 24.0+35.4 | 1.8 | 5.4 | $2\cdot10^{-4}$ |
| | 990516 | (250.0+ 4.1) | 990226.1 | 24.5+38.9 | 1.8 | 8.5 | 3.1 |
| | | | 990318 | 15.0+34.1 | 3.6 | 5.9 | |
| | | | 990408 | 25.6+23.7 | 10.2 | 8.3 | |
| | | | 990516.3 | 15.1+23.9 | 1.8 | 8.5 | |
| 4. | 990706- | 25.4-28.0 | 990706.1 | 33.1- 35.0 | 8.5 | 9.5 | $5\cdot10^{-5}$ |
| | 990904 | (305.4-18.5) | 990713.2 | 17.3- 35.1 | 1.8 | 9.9 | 3.5 |
| | | | 990715.1 | 26.7- 18.1 | 6.5 | 9.9 | |
| | | | 990808 | 29.5- 22.5 | 5.8 | 6.6 | |
| | | | 990904.2 | 14.2  27.9 | 2.5 | 9.9 | |
| 5. | 911106- | 30.4-56.2 | 911106.1 | 8.0- 63.3 | 1.8 | 13.2 | $6\cdot10^{-5}$ |
| | 920411 | (335.7-23.5) | 911128 | 53.8- 53.7 | 6.1 | 13.6 | 3.4 |
| | | | 911224.3 | 44.8- 60.1 | 5.5 | 8.5 | |
| | | | 920127 | 17.8- 45.8 | 8.7 | 13.0 | |
| | | | 920209.1 | 45.0- 46.8 | 8.4 | 13.0 | |
| | | | 920315.1 | 28.5- 42.2 | 1.7 | 14.1 | |
| | | | 920323 | 10.1- 59.2 | 2.4 | 11.2 | |
| | | | 920330 | 10.6- 66.9 | 7.5 | 14.1 | |
| | | | 920407 | 55.9- 56.2 | 2.6 | 14.1 | |
| | | | 920411.2 | 22.6- 60.7 | 9.0 | 6.1 | |
| 6. | 940803- | 30.6+ 7.6 | 940803.2 | 23.6+ 3.0 | 2.0 | 8.4 | $9\cdot10^{-7}$ |
| | 940905 | (275.2+ 1.5) | 940817 | 35.1+ 3.4 | 1.6 | 6.2 | 4.5 |
| | | | 940827.1 | 28.1- 0.5 | 5.1 | 8.4 | |
| | | | 940831 | 34.7+15.0 | 5.3 | 8.4 | |
| | | | 940905 | 31.8+ 1.2 | 4.2 | 6.5 | |
| 7. | 981201- | 39.0-39.3 | 981201 | 36.4- 43.7 | 4.4 | 4.9 | $2\cdot10^{-4}$ |
| | 990303 | (320.9-12.5) | 981211 | 49.7- 48.2 | 1.7 | 11.8 | 3.1 |
| | | | 990117.2 | 25.5- 38.8 | 1.9 | 10.5 | |
| | | | 990126.2 | 24.3- 43.6 | 2.5 | 11.8 | |
| | | | 990202.1 | 47.7- 29.9 | 3.6 | 11.8 | |
| | | | 990303.1 | 32.6- 38.1 | 7.0 | 5.1 | |
| 8. | 921021- | 43.5+55.2 | 921021 | 58.0+48.4 | 2.3 | 11.2 | $6\cdot10^{-5}$ |
| | 930112 | (232.1+28.0) | 921023.1 | 24.5+60.0 | 2.2 | 11.2 | 3.4 |
| | | | 921203 | 32.6+52.5 | 1.8 | 7.0 | |
| | | | 921230.3 | 55.9+64.2 | 4.4 | 10.9 | |
| | | | 930110.2 | 46.7+48.7 | 2.4 | 6.8 | |
| | | | 930112.1 | 42.0+60.4 | 2.0 | 5.3 | |
| 9. | 961025- | 48.6+17.6 | 961025 | 52.1+11.0 | 3.5 | 7.4 | $6\cdot10^{-7}$ |
| | 970302 | (273.5+21.5) | 961102 | 54.4+21.7 | 1.7 | 6.9 | 4.5 |
| | | | 961105 | 41.5+14.5 | 2.4 | 7.5 | |
| | | | 961228.2 | 52.3+24.3 | 4.5 | 7.5 | |
| | | | 970128 | 52.3+17.3 | 2.7 | 3.6 | |
| | | | 970302.2 | 49.1+15.3 | 3.3 | 2.3 | |
| 10. | 941017- | 55.0 - 0.3 | 941017.1 | 50.5- 11.7 | 1.6 | 12.3 | $2\cdot10^{-5}$ |
| | 950117 | (293.8+19.6) | 941110.2 | 54.5- 11.1 | 3.9 | 10.8 | 3.7 |
| | | | 941126.2 | 63.3- 9.4 | 7.6 | 12.3 | |
| | | | 941202 | 54.9+10.8 | 6.1 | 11.1 | |
| | | | 950103 | 45.3+ 6.9 | 10.2 | 12.0 | |
| | | | 950114.2 | 57.0+10.1 | 9.3 | 10.6 | |
| | | | 950117.1 | 60.1+10.9 | 2.0 | 12.3 | |
| 11. | 931001- | 62.4+38.9 | 931001 | 70.8+37.2 | 6.8 | 6.8 | $8\cdot10^{-6}$ |
| | 931223 | (253.4+38.9) | 931030.3 | 56.7+33.8 | 6.1 | 6.9 | 4.0 |
| | | | 931204.1 | 61.3+41.3 | 1.6 | 2.5 | |
| | | | 931218 | 65.7+45.3 | 2.1 | 6.9 | |
| | | | 931223 | 65.8+36.5 | 1.8 | 3.6 | |
| 12. | 970427-<br>970817 | 65.8+22.0<br>(275.3+38.3) | 970427.2 | 73.7+13.1 | 5.0 | 1.6 | $2{\cdot}10^{-6}$<br>4.3 |
| | | | 970430 | 70.6+34.7 | 10.5 | 13.4 | |
| | | | 970525 | 57.3+31.3 | 2.7 | 12.0 | |
| | | | 970612.2 | 75.3+13.4 | 1.7 | 12.5 | |
| | | | 970613.3 | 51.9+19.0 | 3.4 | 13.3 | |
| | | | 970701 | 59.0+22.2 | 7.4 | 6.3 | |
| | | | 970709.1 | 56.7+30.0 | 5.4 | 11.4 | |
| | | | 970725.1 | 79.4+17.9 | 5.6 | 13.4 | |
| | | | 970817.1 | 64.8+19.9 | 2.4 | 2.4 | |
| 13. | 910601-<br>920420 | 70.4 -13.3<br>(314.6+24.9) | 910601 | 73.9- 5.8 | 1.6 | 8.3 | $5{\cdot}10^{-5}$<br>3.5 |
| | | | 910630 | 73.5- 0.5 | 1.8 | 13.2 | |
| | | | 910730.3 | 63.5- 17.8 | 2.7 | 8.0 | |
| | | | 910803 | 66.6- 10.8 | 1.9 | 4.5 | |
| | | | 911016.1 | 61.5- 23.5 | 3.9 | 13.3 | |
| | | | 911025.2 | 80.6- 7.0 | 3.9 | 11.8 | |
| | | | 911110 | 57.2- 13.3 | 2.4 | 12.8 | |
| | | | 911124 | 83.3- 9.4 | 3.5 | 13.2 | |
| | | | 911207 | 70.3- 6.0 | 2.3 | 7.3 | |
| | | | 920210.3 | 75.7- 24.2 | 2.4 | 12.0 | |
| | | | 920404 | 75.4- 20.4 | 1.7 | 8.6 | |
| | | | 920408.3 | 73.5- 2.5 | 4.1 | 11.2 | |
| | | | 920420 | 64.8- 25.0 | 3.8 | 12.9 | |
| 14. | 990707-<br>991201 | 73.7-32.5<br>(331.3+14.2) | 990707.1 | 79.4- 22.4 | 8.7 | 11.3 | $2{\cdot}10^{-4}$<br>3.1 |
| | | | 990807 | 84.5- 31.9 | 2.3 | 9.2 | |
| | | | 990810.2 | 57.9- 35.9 | 6.6 | 13.5 | |
| | | | 990917 | 69.3- 21.2 | 2.5 | 11.9 | |
| | | | 990925.3 | 82.4- 42.0 | 3.7 | 11.7 | |
| | | | 991005.2 | 76.9- 45.5 | 4.0 | 13.2 | |
| | | | 991103 | 66.5- 24.3 | 3.2 | 10.4 | |
| | | | 991201 | 80.6- 20.4 | 4.2 | 13.5 | |
| 15. | 930724-<br>930922 | 85.7+18.8<br>(287.8+54.7) | 930724.1 | 89.3+19.5 | 4.1 | 3.5 | $4{\cdot}10^{-6}$<br>4.1 |
| | | | 930801 | 78.4+12.8 | 2.2 | 9.2 | |
| | | | 930805.1 | 88.8+22.2 | 7.4 | 4.5 | |
| | | | 930913.1 | 77.3+11.1 | 3.9 | 11.2 | |
| | | | 930916.2 | 95.5+25.4 | 1.6 | 11.2 | |
| | | | 930922.2 | 84.2+28.6 | 1.7 | 9.9 | |
| 16. | 970713-<br>971024 | 85.7+20.6<br>(285.3+55.4) | 970713.2 | 87.6+27.2 | 4.0 | 6.8 | $4{\cdot}10^{-6}$<br>4.1 |
| | | | 970725.1 | 79.4+17.9 | 5.6 | 6.5 | |
| | | | 970828 | 88.2+28.2 | 0.05 | 8.0 | |
| | | | 970919.3 | 92.5+15.7 | 3.1 | 8.0 | |
| | | | 971023.2 | 82.3+17.9 | 4.8 | 4.2 | |
| | | | 971024.1 | 77.9+24.1 | 3.0 | 8.0 | |
| 17. | 970817-<br>971223 | 86.1+49.2<br>(234.8+54.0) | 970817.2 | 78.1+43.7 | 9.1 | 7.8 | $4{\cdot}10^{-7}$<br>4.6 |
| | | | 970906 | 98.1+53.9 | 3.0 | 8.8 | |
| | | | 970915 | 94.7+43.9 | 6.4 | 7.9 | |
| | | | 970923 | 80.9+53.4 | 4.9 | 5.3 | |
| | | | 971006 | 81.3+41.0 | 1.7 | 8.9 | |
| | | | 971110 | 78.8+46.5 | 1.7 | 5.6 | |
| | | | 971209 | 94.6+42.7 | 1.8 | 8.7 | |
| | | | 971223 | 87.7+58.0 | 2.7 | 8.8 | |
| 18. | 961030-<br>970110 | 86.4+17.3<br>(290.8+54.8) | 961030 | 97.2+27.2 | 5.4 | 14.0 | $2{\cdot}10^{-4}$<br>3.1 |
| | | | 961211.1 | 84.4+22.5 | 1.9 | 5.6 | |
| | | | 961211.2 | 75.2+ 8.5 | 3.1 | 14.0 | |
| | | | 970105 | 95.9+ 7.2 | 4.7 | 13.7 | |
| | | | 970110 | 83.7+23.1 | 8.2 | 6.3 | |
| 19. | 920628-<br>930114 | 90.1+12.4<br>(301.9+55.9) | 920628.1 | 99.0+19.8 | 9.7 | 11.2 | $6{\cdot}10^{-5}$<br>3.4 |
| | | | 920715 | 84.0+19.9 | 5.9 | 9.5 | |
| | | | 920804.2 | 81.1+18.5 | 2.4 | 10.6 | |
| | | | 920811.1 | 89.1+24.5 | 9.8 | 12.2 | |
| | | | 921011 | 97.5+ 2.7 | 6.8 | 12.1 | |
| | | | 921111.1 | 93.3+13.1 | 2.1 | 3.2 | |
| | | | 921206.1 | 98.9+10.7 | 3.2 | 8.8 | |
| | | | 921214 | 81.6+ 3.7 | 5.9 | 12.1 | |
| | | | 930114.1 | 90.1+22.9 | 5.4 | 10.5 | |
| 20. | 940410-<br>940704 | 96.3+58.6<br>(215.7+53.5) | 940410.1 | 83.6+52.2 | 7.8 | 9.6 | $4{\cdot}10^{-5}$<br>3.6 |
| | | | 940413.2 | 106.6+50.2 | 3.5 | 10.3 | |
| | | | 940512.2 | 89.2+49.4 | 3.1 | 10.1 | |
| | | | 940529.4 | 103.4+54.7 | 1.7 | 5.5 | |
| | | | 940623.2 | 87.0+68.1 | 3.9 | 10.3 | |
| | | | 940704.2 | 90.7+64.7 | 1.7 | 6.6 | |
| 21. | 940209-<br>940428 | 96.4-57.5<br>(358.4+ 2.4) | 940209 | 106.7- 47.1 | 4.9 | 12.1 | $2{\cdot}10^{-4}$<br>3.1 |
| | | | 940214.2 | 81.4- 67.6 | 9.7 | 12.1 | |
| | | | 940226.3 | 85.1- 49.5 | 9.3 | 10.4 | |
| | | | 940314 | 85.3- 59.6 | 1.9 | 6.2 | |
| | | | 940423 | 109.5- 66.4 | 2.9 | 10.8 | |
| | | | 940428 | 75.4- 54.9 | 3.0 | 11.9 | |
| 22. | 000122-<br>000615 | 99.3+30.0<br>(269.5+69.1) | 000122 | 87.6+37.6 | 9.7 | 12.3 | $1{\cdot}10^{-4}$<br>3.3 |
| | | | 000207 | 90.3+19.7 | 1.7 | 13.1 | |
| | | | 000228 | 99.5+43.2 | 4.2 | 13.2 | |
| | | | 000303 | 91.5+27.0 | 3.0 | 7.5 | |
| | | | 000401 | 112.9+24.6 | 3.4 | 13.2 | |
| | | | 000409 | 112.9+29.7 | 5.3 | 11.8 | |
| | | | 000424.1 | 107.6+40.2 | 2.1 | 12.2 | |
| | | | 000615.1 | 109.5+38.7 | 0.0 | 12.1 | |
| 23. | 960201-<br>960704 | 99.8-56.0<br>(359.6+ 4.5) | 960201 | 102.1- 46.0 | 1.6 | 10.1 | $5{\cdot}10^{-5}$<br>3.5 |
| | | | 960214 | 75.9- 58.9 | 5.2 | 13.1 | |
| | | | 960216 | 99.9- 43.0 | 1.7 | 13.1 | |
| | | | 960229.2 | 101.3- 48.0 | 3.0 | 8.1 | |
| | | | 960415.3 | 108.5- 46.6 | 2.6 | 10.9 | |
| | | | 960621.1 | 99.0- 49.1 | 4.0 | 7.0 | |
| | | | 960623.4 | 94.6- 59.7 | 2.9 | 4.6 | |
| | | | 960703.1 | 99.5- 69.1 | 1.9 | 13.1 | |
| | | | 960704 | 85.4- 65.8 | 6.2 | 12.0 | |
| 24. | 980530-<br>980808 | 100.9+68.3<br>(204.4+46.7) | 980530.2 | 121.5+59.8 | 4.8 | 12.3 | $3{\cdot}10^{-5}$<br>3.6 |
| | | | 980609.1 | 101.2+79.2 | 5.9 | 10.9 | |
| | | | 980616 | 66.1+70.4 | 1.9 | 12.3 | |
| | | | 980617.3 | 84.5+58.5 | 5.8 | 12.2 | |
| | | | 980718.2 | 117.6+67.6 | 4.2 | 6.3 | |
| | | | 980808.2 | 115.3+71.9 | 3.4 | 6.1 | |
| 25. | 931015-<br>940217 | 112.2+ 3.7<br>(347.1+64.7) | 931015 | 107.4+ 3.4 | 3.0 | 4.8 | $3{\cdot}10^{-6}$<br>4.2 |
| | | | 931023 | 120.5+11.5 | 1.9 | 11.3 | |
| | | | 931030.2 | 100.5+ 3.7 | 1.7 | 11.7 | |
| | | | 931031 | 102.9+ 7.5 | 1.6 | 10.0 | |
| | | | 931205.2 | 124.0+ 3.6 | 2.7 | 11.8 | |
| | | | 931217 | 119.8+12.7 | 4.1 | 11.7 | |
| | | | 940127 | 110.8+ 9.2 | 8.4 | 5.7 | |
| | | | 940213 | 112.8+ 2.7 | 5.0 | 1.2 | |
| | | | 940217.1 | 111.2- 7.7 | 6.5 | 11.5 | |
| 26. | 970312-<br>970511 | 113.3-30.8<br>(2.5+31.4) | 970312 | 105.9- 26.6 | 3.6 | 7.7 | $7{\cdot}10^{-5}$<br>3.4 |
| | | | 970328.1 | 109.1- 42.1 | 2.7 | 11.8 | |
| | | | 970402.2 | 120.0- 20.7 | 3.9 | 11.8 | |
| | | | 970409 | 124.9- 26.5 | 1.8 | 11.0 | |
| | | | 970414.2 | 105.2- 22.5 | 7.7 | 11.0 | |
| | | | 970511 | 107.3- 41.7 | 2.2 | 11.9 | |
| 27. | 971219-<br>980301 | 114.4-21.3<br>(2.3+40.9) | 971219 | 107.9- 26.2 | 3.6 | 7.7 | $1{\cdot}10^{-4}$<br>3.3 |
| | | | 980124.2 | 119.9- 17.3 | 4.7 | 6.5 | |
| | | | 980125.2 | 103.4- 23.7 | 1.9 | 10.4 | |
| | | | 980206 | 124.3- 16.7 | 4.1 | 10.4 | |
| | | | 980301.4 | 115.0- 14.5 | 2.2 | 6.9 | |
| 28. | 920913-<br>930213 | 117.7- 12.5<br>(4.8+49.4) | 920913 | 110.4- 7.2 | 3.4 | 8.9 | $1{\cdot}10^{-5}$<br>3.9 |
| | | | 920917 | 124.0-15.3 | 2.5 | 6.7 | |
| | | | 921118 | 108.6-10.1 | 1.9 | 9.2 | |
| | | | 921208 | 119.9- 7.0 | 2.9 | 5.9 | |
| | | | 921211 | 126.8-15.2 | 3.5 | 9.2 | |
| | | | 930126 | 114.1-11.7 | 4.0 | 3.6 | |
| | | | 930213.1 | 123.7-13.4 | 3.4 | 5.9 | |
| 29. | 951111-<br>960601 | 119.2+ 3.2<br>(5.7+64.6) | 951111 | 129.4+ 7.6 | 5.9 | 11.0 | $4{\cdot}10^{-5}$<br>3.6 |
| | | | 951119.1 | 122.5+12.1 | 7.1 | 9.5 | |
| | | | 951208.1 | 116.2+ 8.3 | 1.7 | 5.9 | |
| | | | 960209 | 117.1+ 4.2 | 9.5 | 2.3 | |
| | | | 960321 | 120.9+ 4.3 | 1.6 | 2.0 | |
| | | | 960425.2 | 121.2- 6.3 | 2.2 | 9.7 | |
| | | | 960522 | 109.7- 2.4 | 4.3 | 11.0 | |
| | | | 960601.2 | 119.8+ 2.1 | 4.2 | 1.3 | |
| 30. | 940120-<br>940515 | 122.3+29.8<br>(205.1+87.3) | 940120.2 | 127.1+30.5 | 10.2 | 4.2 | $3{\cdot}10^{-5}$<br>3.6 |
| | | | 940210 | 129.0+33.0 | 1.7 | 6.6 | |
| | | | 940318 | 127.8+34.8 | 8.6 | 6.8 | |
| | | | 940425.3 | 115.1+27.2 | 4.0 | 6.8 | |
| | | | 940515.2 | 129.7+27.9 | 2.9 | 6.8 | |
| 31. | 911004-<br>911123 | 122.5+ 8.6<br>(11.8+71.6) | 911004 | 112.6+ 1.9 | 3.1 | 11.9 | $3{\cdot}10^{-5}$<br>3.6 |
| | | | 911022 | 117.8+ 7.3 | 7.6 | 4.8 | |
| | | | 911027.2 | 133.8+ 7.4 | 3.6 | 11.2 | |
| | | | 911119.2 | 120.4+ 3.3 | 7.9 | 5.7 | |
| | | | 911123.1 | 132.3+15.6 | 3.4 | 11.9 | |
| 32. | 000126-<br>000330 | 123.1-27.0<br>(13.1+36.1) | 000126.1 | 111.8- 26.3 | 4.7 | 10.1 | $3{\cdot}10^{-5}$<br>3.6 |
| | | | 000205 | 118.4- 32.5 | 1.9 | 6.9 | |
| | | | 000212 | 125.9- 27.2 | 1.7 | 2.5 | |
| | | | 000217.1 | 126.3- 26.3 | 2.1 | 2.9 | |
| | | | 000317 | 136.7- 28.7 | 3.2 | 12.1 | |
| | | | 000330.2 | 110.8- 22.2 | 5.1 | 12.1 | |
| 33. | 970327 | 134.4-23.3 | 970327.2 | 135.3- 21.5 | 5.7 | 2.0 | $2\cdot10^{-5}$ |
| | 970725 | (26.4+38.5) | 970402.3 | 120.0- 20.7 | 3.9 | 13.6 | 3.7 |
| | | | 970409 | 124.9- 26.5 | 1.8 | 9.2 | |
| | | | 970429.2 | 146.0- 31.9 | 1.9 | 13.4 | |
| | | | 970521 | 136.1- 9.7 | 5.8 | 13.7 | |
| | | | 970616.1 | 141.2- 27.1 | 2.5 | 7.2 | |
| | | | 970629 | 141.6- 35.4 | 2.4 | 13.6 | |
| | | | 970714.1 | 122.3- 30.5 | 3.7 | 12.9 | |
| | | | 970725.2 | 123.0- 26.7 | 3.3 | 10.9 | |
| 34. | 930530- | 135.8-28.3 | 930530.1 | 125.7- 28.4 | 7.5 | 8.9 | $4\cdot10^{-5}$ |
| | 930914 | (26.5+33.3) | 930612.4 | 146.4- 27.7 | 6.6 | 9.4 | 3.6 |
| | | | 930710 | 134.0- 21.5 | 2.5 | 7.0 | |
| | | | 930720.1 | 132.0- 37.0 | 1.7 | 9.3 | |
| | | | 930909 | 125.6- 26.0 | 2.3 | 9.4 | |
| | | | 930914 | 142.4- 25.3 | 2.2 | 6.6 | |
| 35. | 920731- | 137.4+39.7 | 920731 | 120.7+37.9 | 9.1 | 13.2 | $2\cdot10^{-5}$ |
| | 930204 | (152.9+72.6) | 920801 | 139.5+43.5 | 2.0 | 4.1 | 3.7 |
| | | | 920925.2 | 140.0+45.2 | 2.1 | 5.8 | |
| | | | 921029.2 | 152.1+35.4 | 1.9 | 12.4 | |
| | | | 921030.2 | 149.2+40.7 | 6.0 | 9.0 | |
| | | | 921202 | 124.7+49.2 | 4.8 | 13.1 | |
| | | | 921209.1 | 138.7+53.2 | 1.7 | 13.5 | |
| | | | 930201.2 | 123.4+32.4 | 6.3 | 13.5 | |
| | | | 930203.2 | 141.8+41.6 | 2.7 | 3.9 | |
| | | | 930204.1 | 151.4+32.3 | 5.9 | 13.5 | |
| 36. | 991104- | 140.1 - 1.8 | 991104 | 144.7- 0.8 | 1.8 | 4.6 | $1\cdot10^{-5}$ |
| | 000331 | (45.3+56.5) | 991106.2 | 135.5+ 0.4 | 1.9 | 5.1 | 3.9 |
| | | | 000201.2 | 133.0- 4.5 | 5.9 | 7.6 | |
| | | | 000222 | 141.8+ 3.9 | 5.8 | 6.0 | |
| | | | 000302.1 | 147.5+ 0.3 | 1.7 | 7.7 | |
| | | | 000331.3 | 132.4- 1.7 | 4.0 | 7.7 | |
| 37. | 970503- | 142.1+34.2 | 970503 | 133.4+45.1 | 9.7 | 12.8 | $6\cdot10^{-5}$ |
| | 970820 | (131.2+72.1) | 970507 | 130.0+25.4 | 8.3 | 13.7 | 3.4 |
| | | | 970508 | 135.1+26.9 | 0.1 | 9.5 | |
| | | | 970612.3 | 154.6+30.6 | 3.8 | 11.1 | |
| | | | 970713.3 | 158.8+35.4 | 4.2 | 13.7 | |
| | | | 970803 | 153.8+44.6 | 2.7 | 13.7 | |
| | | | 970820 | 145.6+22.5 | 3.3 | 12.1 | |
| 38. | 990308- | 146.2+ 8.1 | 990308.1 | 135.0+ 9.7 | 5.3 | 11.2 | $5\cdot10^{-5}$ |
| | 990523 | (66.2+60.9) | 990308.3 | 139.0+14.0 | 4.2 | 9.2 | 3.5 |
| | | | 990315.1 | 146.9+11.0 | 4.2 | 3.0 | |
| | | | 990329 | 142.2+14.9 | 2.1 | 7.9 | |
| | | | 990516.1 | 156.1+13.4 | 3.5 | 11.1 | |
| | | | 990523.3 | 145.2 - 3.0 | 5.5 | 11.1 | |
| 39. | 000126- | 146.7+32.3 | 000126.3 | 150.2+38.8 | 8.6 | 7.0 | $5\cdot10^{-6}$ |
| | 000730 | (122.3+68.8) | 000221 | 136.2+31.3 | 1.7 | 9.0 | 4.1 |
| | | | 000408.2 | 147.5+38.6 | 0.1 | 6.3 | |
| | | | 000415.2 | 144.7+36.3 | 1.8 | 4.4 | |
| | | | 000517 | 137.9+27.5 | 1.9 | 9.0 | |
| | | | 000620 | 146.5+29.3 | 0.1 | 3.0 | |
| | | | 000730.2 | 156.9+35.3 | 0.1 | 9.0 | |
| 40. | 951112- | 148.6+ 9.9 | 951112.1 | 149.7+22.9 | 8.7 | 13.0 | $2\cdot10^{-4}$ |
| | 960124 | ( 71.6+60.3) | 951117.2 | 155.8+ 5.8 | 7.1 | 8.2 | 3.1 |
| | | | 951124.1 | 155.7+ 4.9 | 2.0 | 8.6 | |
| | | | 960119.1 | 157.6+13.0 | 1.8 | 9.4 | |
| | | | 960124.1 | 144.2- 2.4 | 1.6 | 13.0 | |
| 41. | 950225- | 154.3+ 5.0 | 950225.1 | 158.8- 3.3 | 3.1 | 9.5 | $9\cdot10^{-5}$ |
| | 950419 | (71.9+52.5) | 950227 | 162.1+15.5 | 4.1 | 12.9 | 3.3 |
| | | | 950325.1 | 147.1- 4.7 | 1.7 | 12.1 | |
| | | | 950401.2 | 167.2+ 7.4 | 1.6 | 13.0 | |
| | | | 950419 | 143.6- 2.3 | 2.7 | 12.9 | |
| 42. | 950207- | 154.5-23.6 | 950207.2 | 149.0- 31.7 | 3.1 | 9.4 | $4\cdot10^{-5}$ |
| | 950509 | (46.2+30.8) | 950317.2 | 156.9- 26.1 | 2.0 | 3.3 | 3.6 |
| | | | 950327.1 | 155.4- 14.3 | 2.1 | 9.4 | |
| | | | 950430.1 | 163.5- 19.5 | 2.0 | 9.3 | |
| | | | 950509.2 | 157.3- 31.4 | 3.6 | 8.0 | |
| 43. | 970227- | 157.6+27.0 | 970227.2 | 164.1+14.9 | 7.4 | 13.5 | $2\cdot10^{-6}$ |
| | 970321 | (110.5+59.3) | 970313 | 145.9+36.0 | 4.3 | 13.4 | 4.3 |
| | | | 970317.2 | 146.8+28.4 | 2.6 | 9.7 | |
| | | | 970318 | 160.8+40.1 | 6.9 | 13.4 | |
| | | | 970321 | 163.0+16.3 | 6.0 | 11.8 | |
| 44. | 930926- | 165.4- 9.6 | 930926 | 168.3- 12.9 | 4.0 | 4.2 | $2\cdot10^{-4}$ |
| | 931229 | ( 67.1+34.7) | 931016.2 | 165.1+ 0.1 | 6.3 | 9.8 | 3.1 |
| | | | 931115 | 170.9- 17.8 | 3.2 | 9.7 | |
| | | | 931211 | 171.5- 9.2 | 6.6 | 5.9 | |
| | | | 931229.2 | 156.3- 13.5 | 3.6 | 9.8 | |
| 45. | 921110- | 167.9- 5.7 | 921110.3 | 161.8- 14.5 | 5.1 | 10.6 | $2\cdot10^{-4}$ |
| | 930123 | (72.8+35.2) | 921112.4 | 174.9- 5.4 | 2.6 | 7.0 | 3.1 |
| | | | 930104.1 | 172.7+ 3.9 | 10.2 | 10.7 | |
| | | | 930109 | 177.9- 8.5 | 3.9 | 10.3 | |
| | | | 930123 | 159.0- 10.9 | 3.0 | 10.2 | |
| 46. | 950803- | 169.9+50.5 | 950803 | 159.1+39.5 | 10.2 | 13.4 | $4\cdot10^{-6}$ |
| | 960111 | (148.7+47.6) | 950901.1 | 174.0+38.6 | 3.9 | 12.3 | 4.1 |
| | | | 950921.1 | 167.5+56.3 | 1.8 | 6.0 | |
| | | | 950922.2 | 184.3+61.2 | 2.2 | 13.4 | |
| | | | 950926 | 161.6+62.1 | 3.3 | 12.5 | |
| | | | 951102.2 | 172.7+43.5 | 1.7 | 7.3 | |
| | | | 951107.3 | 182.1+51.6 | 4.3 | 7.7 | |
| | | | 951112.2 | 170.3+39.3 | 3.3 | 11.2 | |
| | | | 960107 | 155.3+42.1 | 5.0 | 13.1 | |
| | | | 960111 | 171.1+43.0 | 1.7 | 7.5 | |
| 47. | 960621- | 171.8+ 9.2 | 960621.2 | 179.4+ 4.6 | 4.6 | 8.9 | $1\cdot10^{-6}$ |
| | 960819 | (90.9+40.5) | 960624.2 | 182.4+11.0 | 2.4 | 10.6 | 4.4 |
| | | | 960715.2 | 161.2+ 7.6 | 3.1 | 10.6 | |
| | | | 960723 | 170.7+12.9 | 2.2 | 3.9 | |
| | | | 960808 | 172.6+ 3.4 | 1.7 | 5.8 | |
| | | | 960812 | 175.3+12.5 | 2.9 | 4.8 | |
| | | | 960819 | 161.3+11.6 | 4.8 | 10.6 | |
| 48. | 920210- | 172.5+50.3 | 920210.2 | 168.2+51.4 | 1.7 | 2.9 | $2\cdot10^{-5}$ |
| | 920503 | (148.4+46.1) | 920218.2 | 164.3+43.7 | 3.7 | 8.7 | 3.7 |
| | | | 920303.1 | 165.9+58.1 | 6.4 | 8.7 | |
| | | | 920408.2 | 181.2+44.0 | 1.7 | 8.6 | |
| | | | 920502.2 | 171.5+54.8 | 1.7 | 4.5 | |
| | | | 920503.1 | 179.0+49.8 | 3.5 | 4.2 | |
| 49. | 930608- | 174.8+16.6 | 930608.2 | 176.0+18.9 | 10.4 | 2.6 | $1\cdot10^{-4}$ |
| | 930731 | (101.5+41.3) | 930612.2 | 174.7+ 4.9 | 2.5 | 11.7 | 3.3 |
| | | | 930720.2 | 167.2+14.9 | 3.2 | 7.5 | |
| | | | 930731.1 | 173.6+28.2 | 2.7 | 11.7 | |
| | | | 930731.2 | 178.6+15.7 | 1.7 | 3.7 | |
| 50. | 911217- | 176.5-53.4 | 911217.2 | 155.1- 58.8 | 6.0 | 13.0 | $1\cdot10^{-4}$ |
| | 920214 | (41.6- 3.3) | 911224.2 | 191.2- 55.3 | 4.9 | 8.8 | 3.3 |
| | | | 920130.2 | 189.4- 43.5 | 3.7 | 13.0 | |
| | | | 920205 | 193.7- 62.9 | 8.7 | 13.1 | |
| | | | 920214 | 164.0- 47.0 | 4.6 | 10.2 | |
| 51. | 970523- | 177.3+19.0 | 970523.2 | 174.3+24.1 | 2.8 | 5.8 | $2\cdot10^{-5}$ |
| | 970910 | (105.8+39.9) | 970613.1 | 188.9+15.7 | 4.5 | 11.5 | 3.7 |
| | | | 970627.2 | 177.8+30.7 | 1.8 | 11.7 | |
| | | | 970801 | 185.5+11.6 | 1.8 | 10.8 | |
| | | | 970816 | 168.0+11.5 | 1.6 | 11.7 | |
| | | | 970824.3 | 182.9+25.2 | 7.2 | 8.1 | |
| | | | 970910.2 | 182.9+12.2 | 6.0 | 8.7 | |
| 52. | 000108- | 183.4+40.8 | 000108.2 | 176.1+30.6 | 7.5 | 11.8 | $3\cdot10^{-4}$ |
| | 000314 | (134.8+39) | 000110 | 174.7+50.9 | 5.6 | 11.8 | ~3.0 |
| | | | 000114.2 | 181.1+30.8 | 8.1 | 10.2 | |
| | | | 000226.1 | 197.3+47.0 | 2.3 | 11.8 | |
| | | | 000314 | 167.8+42.5 | 2.3 | 11.8 | |
| 53. | 940823- | 190.7+ 9.3 | 940823.2 | 201.2+ 4.0 | 4.6 | 11.7 | $4\cdot10^{-6}$ |
| | 941114 | (101.1+24.0) | 940916 | 192.6+20.8 | 3.3 | 11.7 | 4.1 |
| | | | 940921.1 | 194.2+18.6 | 2.1 | 9.9 | |
| | | | 941003.2 | 199.2+ 9.9 | 4.3 | 8.4 | |
| | | | 941026.2 | 184.4+19.4 | 1.8 | 11.8 | |
| | | | 941111 | 181.3+10.2 | 9.7 | 9.3 | |
| | | | 941114 | 179.0+11.7 | 3.1 | 11.8 | |
| 54. | 940902- | 192.5-43.1 | 940902.1 | 192.3- 55.4 | 4.6 | 12.3 | $3\cdot10^{-6}$ |
| | 941018 | (55.7- 4.8) | 940902.2 | 195.6- 44.4 | 5.0 | 2.6 | 4.2 |
| | | | 940909 | 182.5- 38.2 | 6.7 | 9.0 | |
| | | | 940918 | 190.6- 30.9 | 2.8 | 12.3 | |
| | | | 941018.1 | 206.7- 48.9 | 1.8 | 11.4 | |
| | | | 941018.3 | 190.1- 43.9 | 8.2 | 1.9 | |
| 55. | 921003- | 198.9- 9.4 | 921003.2 | 191.8- 8.2 | 5.5 | 7.1 | $4\cdot10^{-5}$ |
| | 930305 | (88.0+ 7.9) | 921013 | 204.3- 13.1 | 1.9 | 6.5 | 3.6 |
| | | | 921030.1 | 199.0- 11.5 | 2.4 | 2.1 | |
| | | | 921217 | 187.8- 2.0 | 2.9 | 13.3 | |
| | | | 930110.1 | 210.0- 1.9 | 3.7 | 13.4 | |
| | | | 930131.2 | 204.4- 21.6 | 5.5 | 13.3 | |
| | | | 930302 | 202.3- 8.6 | 1.7 | 3.4 | |
| | | | 930304 | 187.1- 12.2 | 4.4 | 12.0 | |
| | | | 930305.2 | 199.6- 0.2 | 4.3 | 9.2 | |
| 56. | 930607- | 203.3+71.1 | 930607 | 186.8+63.6 | 3.1 | 9.8 | $6\cdot10^{-6}$ |
| | 931013 | (171.0+28.7) | 930614.1 | 215.5+67.5 | 2.0 | 5.7 | 4.0 |
| | | | 930812 | 208.7+66.5 | 7.9 | 5.1 | |
| | | | 930826 | 199.8+74.2 | 5.0 | 3.2 | |
| | | | 930828 | 171.7+73.3 | 3.1 | 9.8 | |
| | | | 931008.3 | 190.1+65.5 | 1.8 | 7.3 | |
| | | | 931013.1 | 228.2+66.9 | 2.7 | 9.8 | |
| 57. | 971024- | 203.8+ 8.3 | 971024.2 | 190.5+12.8 | 4.7 | 13.9 | $2\cdot10^{-5}$ |
| | 980207 | (106.3+11.8) | 971029.1 | 190.0+ 9.5 | 1.6 | 13.7 | 3.7 |
| | | | 971206.1 | 216.1+15.2 | 6.3 | 13.9 | |
| | | | 971207.2 | 206.3 - 5.4 | 2.1 | 13.9 | |
| | | | 971218.2 | 203.5+19.1 | 5.4 | 10.8 | |
| | | | 980106 | 203.3+ 9.4 | 3.4 | 1.2 | |
| | | | 980125.1 | 209.0 - 4.4 | 3.7 | 13.7 | |
| | | | 980207.1 | 207.0+ 1.2 | 2.2 | 7.8 | |
| 58. | 980327- | 204.8+ 5.6 | 980327 | 193.0+11.0 | 5.7 | 12.8 | $3\cdot10^{-6}$ |
| | 980527 | (104.0+ 9.8) | 980407 | 212.9+ 2.4 | 8.1 | 8.7 | 4.2 |
| | | | 980413 | 201.1 - 4.3 | 5.1 | 10.5 | |
| | | | 980416 | 209.6 - 6.3 | 7.9 | 12.8 | |
| | | | 980424 | 203.9 - 6.1 | 3.5 | 11.7 | |
| | | | 980503 | 204.9+18.3 | 6.2 | 12.7 | |
| | | | 980527.2 | 202.3+11.4 | 8.5 | 6.3 | |
| 59. | 920423- | 210.6+57.9 | 920423.2 | 212.2+54.4 | 5.7 | 3.6 | $3\cdot10^{-7}$ |
| | 921230 | (157.2+23.8) | 920524 | 200.7+58.9 | 2.5 | 5.3 | 4.7 |
| | | | 920530 | 224.5+60.8 | 2.3 | 7.6 | |
| | | | 920619.2 | 212.3+55.7 | 5.6 | 2.4 | |
| | | | 920721.1 | 206.9+54.4 | 10.4 | 4.1 | |
| | | | 920920 | 214.1+50.8 | 5.0 | 7.4 | |
| | | | 921110.2 | 221.0+54.0 | 3.2 | 7.0 | |
| | | | 921216 | 215.3+61.2 | 4.6 | 4.1 | |
| | | | 921230.2 | 200.5+52.8 | 2.3 | 7.7 | |
| 60. | 961115- | 223.5-49.7 | 961115 | 230.0- 50.1 | 2.5 | 4.2 | $1\cdot10^{-5}$ |
| | 970226 | (58.5-27.0) | 961216 | 219.8- 38.3 | 3.7 | 11.7 | 3.9 |
| | | | 970201.1 | 210.9- 50.6 | 1.6 | 8.1 | |
| | | | 970202 | 224.6- 57.6 | 1.6 | 8.0 | |
| | | | 970224 | 239.3- 45.1 | 2.1 | 11.6 | |
| | | | 970226.2 | 218.8- 61.0 | 2.1 | 11.6 | |
| 61. | 910828- | 225.9-50.9 | 910828 | 230.9- 63.1 | 5.4 | 12.5 | $7\cdot10^{-4}$ |
| | 920110 | (57.3-28.6) | 910923 | 210.5- 52.1 | 4.3 | 7.4 | 3.0 |
| | | | 910927.2 | 244.7- 56.7 | 1.6 | 12.5 | |
| | | | 911204 | 219.6- 39.2 | 2.5 | 12.5 | |
| | | | 911209.2 | 229.2- 60.8 | 2.0 | 10.1 | |
| | | | 920110.1 | 214.3- 48.4 | 1.7 | 7.8 | |
| 62. | 000418- | 232.4+78.9 | 000418.1 | 264.1+81.1 | 0.1 | 5.8 | $2\cdot10^{-5}$ |
| | 000730 | (180.9+23) | 000421 | 238.7+69.4 | 1.7 | 9.6 | 3.7 |
| | | | 000616 | 210.5+70.1 | 0.1 | 10.4 | |
| | | | 000727 | 243.1+72.3 | 0.1 | 7.1 | |
| | | | 000730.1 | 296.0+82.0 | 0.1 | 10.4 | |
| 63. | 930309- | 234.5-77.9 | 930309.2 | 220.9-72.5 | 7.5 | 6.4 | $2\cdot10^{-4}$ |
| | 930519 | (25.5-31.0) | 930410 | 185.0-77.1 | 2.1 | 10.4 | 3.1 |
| | | | 930420.1 | 223.7-69.4 | 5.0 | 9.0 | |
| | | | 930425.1 | 275.6-80.8 | 1.6 | 7.9 | |
| | | | 930519 | 267.8-71.6 | 4.4 | 10.5 | |
| 64. | 940321- | 234.9+48.7 | 940321 | 245.0+50.7 | 1.8 | 6.8 | $6\cdot10^{-6}$ |
| | 940727 | (154.8+ 7.2) | 940503 | 236.4+55.5 | 1.7 | 6.8 | 4.0 |
| | | | 940524.2 | 231.2+53.5 | 6.0 | 5.3 | |
| | | | 940614 | 234.7+53.7 | 7.0 | 5.0 | |
| | | | 940713 | 225.7+45.7 | 7.0 | 6.9 | |
| | | | 940727 | 242.6+44.5 | 8.7 | 6.8 | |
| 65. | 911109- | 238.1- 5.9 | 911109 | 240.6- 4.4 | 1.6 | 2.9 | $4\cdot10^{-8}$ |
| | 911226 | (109.3-24.9) | 911122 | 239.8- 9.9 | 4.3 | 4.3 | >5.0 |
| | | | 911205 | 236.7- 10.6 | 1.9 | 4.9 | |
| | | | 911208.1 | 236.1- 1.3 | 4.2 | 5.0 | |
| | | | 911226 | 242.9- 6.3 | 7.4 | 4.8 | |
| 66. | 940224- | 238.8+20.9 | 940224 | 237.7+25.5 | 6.5 | 4.7 | $1\cdot10^{-8}$ |
| | 940829 | (133.8-11.6) | 940228.1 | 236.2+15.8 | 1.7 | 5.7 | >5.0 |
| | | | 940228.2 | 234.2+14.9 | 5.6 | 7.4 | |
| | | | 940413.3 | 237.2+24.5 | 3.8 | 3.9 | |
| | | | 940524.1 | 232.9+26.6 | 2.8 | 7.8 | |
| | | | 940604.2 | 236.2+13.8 | 2.2 | 7.5 | |
| | | | 940714.3 | 246.4+17.7 | 2.0 | 7.8 | |
| | | | 940822 | 236.4+23.2 | 3.1 | 3.2 | |
| | | | 940827.2 | 243.5+17.8 | 3.2 | 5.4 | |
| | | | 940829 | 239.6+15.2 | 3.0 | 5.8 | |
| 67. | 910807- | 247.6+52.9 | 910807 | 234.8+47.6 | 1.6 | 9.7 | $3\cdot10^{-4}$ |
| | 920216 | (163.1+ 3.2) | 910816.2 | 262.4+57.8 | 3.7 | 9.7 | 3.0 |
| | | | 910926.1 | 246.8+45.4 | 5.5 | 7.5 | |
| | | | 911126 | 239.6+52.2 | 1.6 | 4.6 | |
| | | | 920110.2 | 244.3+50.5 | 4.5 | 3.2 | |
| | | | 920216.2 | 243.8+46.9 | 1.9 | 6.5 | |
| 68. | 970206- | 248.7+61.9 | 970206 | 234.8+66.4 | 2.0 | 7.5 | $1\cdot10^{-7}$ |
| | 970408 | (170.2+ 9.1) | 970308 | 241.5+60.3 | 6.0 | 3.9 | 4.9 |
| | | | 970329.2 | 248.6+61.5 | 4.9 | 0.4 | |
| | | | 970407.1 | 263.7+59.8 | 4.1 | 7.6 | |
| | | | 970408.1 | 237.1+57.2 | 3.3 | 7.5 | |
| | | | 970408.2 | 248.3+59.2 | 9.1 | 2.7 | |
| 69. | 961027- | 255.2- 43.9 | 961027.1 | 246.8- 43.5 | 5.9 | 6.1 | $7\cdot10^{-10}$ |
| | 961029 | (66.2-48.4) | 961027.2 | 262.8- 41.4 | 6.0 | 6.1 | >5.0 |
| | | | 961029.1 | 262.5- 47.2 | 3.7 | 6.1 | |
| | | | 961029.2 | 257.1- 48.0 | 1.6 | 4.3 | |
| 70. | 940521- | 266.3-13.7 | 940521.1 | 278.2- 13.1 | 4.4 | 11.6 | $1\cdot10^{-4}$ |
| | 940803 | (116.9-53.3) | 940529.2 | 256.7- 20.9 | 2.1 | 11.6 | 3.3 |
| | | | 940604.1 | 264.6- 2.3 | 5.8 | 11.6 | |
| | | | 940717.2 | 269.2- 23.4 | 9.2 | 10.1 | |
| | | | 940731 | 270.4- 12.9 | 5.6 | 4.0 | |
| | | | 940803.1 | 272.3- 19.3 | 4.4 | 8.0 | |
| 71. | 920302- | 271.8-18.9 | 920302.1 | 271.1- 17.0 | 2.6 | 2.0 | $2\cdot10^{-4}$ |
| | 920925 | (112.2-60.3) | 920314.1 | 277.1- 24.9 | 9.0 | 7.8 | 3.1 |
| | | | 920413 | 260.2–15.9 | 3.5 | 11.5 | |
| | | | 920414.2 | 285.4- 23.3 | 2.2 | 13.4 | |
| | | | 920511.1 | 265.2- 7.1 | 3.5 | 13.5 | |
| | | | 920617.1 | 274.6- 25.2 | 2.3 | 6.8 | |
| | | | 920619.1 | 282.7- 10.5 | 2.8 | 13.5 | |
| | | | 920802 | 257.5- 20.8 | 2.2 | 13.6 | |
| | | | 920924 | 282.6- 17.1 | 5.1 | 10.4 | |
| | | | 920925.1 | 274.8- 10.4 | 5.3 | 9.0 | |
| 72. | 940424- | 272.4-57.9 | 940424 | 270.6- 67.4 | 3.7 | 8.4 | $5\cdot10^{-5}$ |
| | 940911 | (39.2-52.5) | 940429.1 | 174.4- 62.2 | 1.6 | 4.4 | 3.5 |
| | | | 940527.1 | 268.4- 68.6 | 4.7 | 9.8 | |
| | | | 940728.1 | 264.8- 58.3 | 1.8 | 7.0 | |
| | | | 940806.4 | 281.5- 64.9 | 1.7 | 5.4 | |
| | | | 940901 | 259.4- 61.5 | 7.9 | 9.3 | |
| | | | 940911 | 286.8- 51.2 | 5.1 | 9.8 | |
| 73. | 990226- | 277.5- 2.7 | 990226.2 | 276.9+ 0.6 | 4.5 | 3.4 | $2\cdot10^{-4}$ |
| | 990523 | (143.3-55.3) | 990305.2 | 279.3+ 6.3 | 3.2 | 9.2 | 3.1 |
| | | | 990328 | 275.3+ 6.3 | 4.6 | 9.3 | |
| | | | 990412 | 276.5- 7.7 | 9.9 | 5.1 | |
| | | | 990523.1 | 278.0- 11.9 | 2.5 | 9.2 | |
| 74. | 941031- | 280.9-46.0 | 941031.2 | 278.2- 54.4 | 3.8 | 8.6 | $3\cdot10^{-8}$ |
| | 950608 | (49.8-64.3) | 941121.2 | 289.5- 48.3 | 2.5 | 6.3 | >5.0 |
| | | | 941128.1 | 295.1- 43.1 | 3.3 | 10.5 | |
| | | | 941229.1 | 275.6- 36.9 | 7.8 | 9.9 | |
| | | | 950105 | 294.9- 53.6 | 2.6 | 11.8 | |
| | | | 950118.2 | 277.1- 56.2 | 3.4 | 10.5 | |
| | | | 950211.2 | 264.6- 43.7 | 2.0 | 11.8 | |
| | | | 950225.2 | 281.7- 31.4 | 8.6 | 11.7 | |
| | | | 950409 | 279.1- 48.5 | 3.5 | 5.7 | |
| | | | 950421 | 273.4- 37.8 | 1.8 | 9.9 | |
| | | | 950602 | 278.5- 52.8 | 1.8 | 7.0 | |
| | | | 950608 | 264.4- 47.9 | 1.8 | 11.4 | |
| 75. | 920502- | 299.6-37.3 | 920502.3 | 298.1- 24.6 | 4.1 | 12.8 | $2\cdot10^{-5}$ |
| | 920606 | (27.1-79.5) | 920509 | 307.7- 33.5 | 4.0 | 7.6 | 3.7 |
| | | | 920517.1 | 299.4- 33.6 | 6.2 | 3.7 | |
| | | | 920525.3 | 301.9- 50.0 | 3.7 | 12.8 | |
| | | | 920606 | 315.6- 39.5 | 8.4 | 12.7 | |
| 76. | 940307- | 304.7+38.8 | 940307 | 302.4+34.7 | 3.9 | 4.5 | $2\cdot10^{-5}$ |
| | 940507 | (194.4-24.1) | 940330.2 | 303.8+29.2 | 2.3 | 9.6 | 3.7 |
| | | | 940413.1 | 318.6+38.0 | 3.5 | 10.9 | |
| | | | 940414.2 | 290.1+35.6 | 1.7 | 12.0 | |
| | | | 940421 | 291.4+41.2 | 2.1 | 10.5 | |
| | | | 940507 | 320.3+39.0 | 8.5 | 12.1 | |
| 77. | 961228- | 313.1-53.1 | 961228.3 | 311.3- 57.5 | 1.9 | 4.5 | $5\cdot10^{-6}$ |
| | 970406 | (359.1-62.9) | 970108 | 317.8- 62.2 | 5.2 | 9.4 | 4.1 |
| | | | 970221 | 316.9- 44.0 | 2.2 | 9.4 | |
| | | | 970302.1 | 309.1- 62.3 | 4.5 | 9.5 | |
| | | | 970326.1 | 310.7- 48.7 | 3.7 | 4.7 | |
| | | | 970406.1 | 319.3- 45.0 | 3.2 | 9.0 | |
| 78. | 980103- | 314.0- 3.9 | 980103.1 | 313.7- 15.9 | 2.5 | 12.1 | $3 \cdot 10^{-5}$ |
| | 980401 | (219.3-64.5) | 980124.3 | 302.6+ 0.2 | 4.2 | 12.1 | 3.6 |
| | | | 980126 | 324.9- 8.8 | 2.6 | 11.9 | |
| | | | 980218.3 | 312.7- 4.8 | 5.2 | 1.6 | |
| | | | 980310.2 | 325.1- 6.4 | 2.0 | 11.4 | |
| | | | 980325.1 | 312.4+ 8.1 | 2.1 | 12.1 | |
| | | | 980401.3 | 305.6+ 0.4 | 2.9 | 9.4 | |
| 79. | 990315- | 314.4-42.6 | 990315.2 | 308.4- 53.8 | 3.1 | 11.9 | $1 \cdot 10^{-6}$ |
| | 990513 | (345.1-72.0) | 990322 | 303.1- 48.8 | 2.4 | 10.0 | 4.4 |
| | | | 990323.3 | 313.8- 37.0 | 1.8 | 5.6 | |
| | | | 990404.2 | 322.5- 32.6 | 7.7 | 11.8 | |
| | | | 990411.1 | 308.2- 31.8 | 1.7 | 11.8 | |
| | | | 990510.1 | 304.1- 49.8 | 9.4 | 10.1 | |
| | | | 990513.2 | 313.5- 39.2 | 2.4 | 3.4 | |
| 80. | 980320- | 315.8+18.8 | 980320 | 329.7+17.0 | 4.9 | 13.3 | $2 \cdot 10^{-4}$ |
| | 980601 | (209.4-42.3) | 980325.1 | 312.4+ 8.1 | 2.1 | 11.2 | 3.1 |
| | | | 980331.2 | 301.8+17.9 | 5.3 | 13.3 | |
| | | | 980426.1 | 317.1+31.6 | 3.4 | 12.9 | |
| | | | 980523 | 307.5+20.3 | 4.3 | 7.9 | |
| | | | 980601 | 320.8+ 6.6 | 4.1 | 13.1 | |
| 81. | 941121- | 316.0-57.2 | 941121.1 | 318.4- 69.6 | 5.1 | 12.4 | $3 \cdot 10^{-5}$ |
| | 950129 | (359.3-58.5) | 941126.1 | 302.2- 54.1 | 4.9 | 8.4 | 3.6 |
| | | | 941228 | 331.5- 63.5 | 2.3 | 9.8 | |
| | | | 950105 | 294.9- 53.6 | 2.6 | 12.4 | |
| | | | 950111.1 | 324.3- 45.8 | 2.0 | 12.5 | |
| | | | 950129.2 | 335.7- 59.0 | 4.5 | 10.5 | |
| 82. | 970930- | 316.4-22.8 | 970930.1 | 309.1- 26.9 | 1.8 | 7.8 | $2 \cdot 10^{-6}$ |
| | 971124 | (266.1-77.0) | 970930.2 | 305.1- 26.7 | 4.0 | 10.9 | 4.3 |
| | | | 971009.2 | 328.2- 21.0 | 1.7 | 11.0 | |
| | | | 971108 | 316.5- 12.3 | 5.2 | 10.5 | |
| | | | 971113 | 308.5- 14.8 | 1.8 | 11.0 | |
| | | | 971124 | 305.8- 25.7 | 9.7 | 10.1 | |
| 83. | 920813- | 318.7+37.2 | 920813 | 317.0+33.1 | 2.4 | 4.4 | $2 \cdot 10^{-4}$ |
| | 930106 | (206.4-24.2) | 921017 | 310.6+45.3 | 6.3 | 10.1 | 3.1 |
| | | | 921102.2 | 329.8+31.9 | 4.9 | 10.6 | |
| | | | 921109 | 322.2+26.7 | 3.3 | 10.9 | |
| | | | 921227 | 317.9+35.7 | 1.9 | 1.7 | |
| | | | 930106.1 | 312.3+47.0 | 2.4 | 10.9 | |
| 84. | 960202- | 318.8-14.1 | 960202 | 312.3- 3.4 | 2.2 | 12.5 | $9 \cdot 10^{-6}$ |
| | 960329 | (244.8-70.1) | 960311 | 306.2- 12.0 | 6.4 | 12.4 | 3.9 |
| | | | 960312 | 322.3- 6.0 | 4.8 | 8.8 | |
| | | | 960316 | 312.9- 9.7 | 2.1 | 7.3 | |
| | | | 960329 | 327.1- 23.9 | 7.9 | 12.5 | |
| 85. | 911223- | 321.1-58.7 | 911223 | 342.6- 54.2 | 3.1 | 12.6 | $5 \cdot 10^{-5}$ |
| | 920329 | (356.0-56.0) | 911227 | 314.8- 46.7 | 1.7 | 12.5 | 3.5 |
| | | | 920105 | 338.4- 51.7 | 2.8 | 12.0 | |
| | | | 920130.1 | 310.8- 65.8 | 1.8 | 8.5 | |
| | | | 920307 | 335.8- 67.3 | 2.0 | 10.8 | |
| | | | 920314.3 | 314.1- 55.5 | 6.7 | 5.0 | |
| | | | 920329 | 296.3- 62.1 | 1.8 | 12.6 | |
| 86. | 990414- | 322.3+ 5.2 | 990414 | 333.0+ 6.1 | 2.0 | 10.7 | $1 \cdot 10^{-4}$ |
| | 990619 | (225.7-52.6) | 990505 | 325.8+ 8.9 | 1.9 | 5.1 | 3.3 |
| | | | 990523.2 | 322.7- 4.3 | 10.5 | 9.5 | |
| | | | 990618 | 313.2- 0.2 | 2.3 | 10.6 | |
| | | | 990619 | 325.9+15.3 | 2.0 | 10.7 | |
| 87. | 990102- | 322.4+31.5 | 990102.2 | 331.0+33.2 | 7.1 | 7.4 | $5 \cdot 10^{-5}$ |
| | 990304 | (211.6-28.3) | 990129.3 | 314.9+33.5 | 7.5 | 6.6 | 3.5 |
| | | | 990213.1 | 313.8+34.2 | 3.9 | 7.7 | |
| | | | 990225.1 | 314.9+27.6 | 2.6 | 7.6 | |
| | | | 990304.2 | 331.1+29.9 | 3.2 | 7.7 | |
| 88. | 000214- | 328.1-16.6 | 000214 | 329.2- 25.2 | 0.1 | 8.7 | $2 \cdot 10^{-5}$ |
| | 000416 | (263.2-64.3) | 000217.2 | 337.1- 17.3 | 4.9 | 8.7 | 3.7 |
| | | | 000326.2 | 330.5- 19.0 | 5.2 | 3.3 | |
| | | | 000408.1 | 319.6- 17.5 | 4.3 | 8.2 | |
| | | | 000416 | 320.8- 18.8 | 0.0 | 7.3 | |
| 89. | 980214- | 336.0-79.4 | 980214.1 | 10.2- 73.5 | 2.9 | 9.7 | $1 \cdot 10^{-4}$ |
| | 980421 | (5.7-35.9) | 980306.1 | 315.5- 71.0 | 1.6 | 9.7 | 3.3 |
| | | | 980315.4 | 337.9- 73.4 | 5.1 | 6.0 | |
| | | | 980319 | 36.2- 87.7 | 3.0 | 9.7 | |
| | | | 980421.2 | 350.6- 71.2 | 4.2 | 9.0 | |
| 90. | 940419-<br>940821 | 336.1-59.5<br>(347.0-50.4) | 940419.2 | 325.5- 66.8 | 1.7 | 8.7 | $5 \cdot 10^{-5}$<br>3.5 |
| | | | 40512.3 | 334.5- 47.3 | 9.3 | 12.3 | |
| | | | 940515.1 | 355.7- 60.8 | 3.6 | 9.8 | |
| | | | 940714.1 | 357.5- 54.8 | 2.3 | 12.4 | |
| | | | 940726 | 313.2- 66.7 | 2.0 | 12.5 | |
| | | | 940730 | 320.5- 50.8 | 2.3 | 12.4 | |
| | | | 940812.1 | 347.7- 58.2 | 2.7 | 6.1 | |
| | | | 940821.1 | 352.1- 54.4 | 4.0 | 10.0 | |
| 91. | 960613-<br>960815 | 340.0-23.4<br>(284.8-56.4) | 960613.1 | 328.3 -22.3 | 3.8 | 10.8 | $4 \cdot 10^{-5}$<br>3.6 |
| | | | 960615.3 | 330.2- 29.7 | 7.6 | 10.8 | |
| | | | 960623.2 | 351.7- 21.8 | 7.7 | 10.9 | |
| | | | 960730.1 | 338.7- 19.8 | 3.2 | 3.8 | |
| | | | 960804.1 | 343.8- 20.2 | 8.5 | 4.8 | |
| | | | 960815 | 334.3- 14.2 | 1.9 | 10.7 | |
| 92. | 930610-<br>930831 | 347.2+54.1<br>(217.1- 0.4) | 930610 | 338.2+55.7 | 8.1 | 5.4 | $2 \cdot 10^{-7}$<br>4.8 |
| | | | 930706.2 | 335.9+57.0 | 2.1 | 7.0 | |
| | | | 930715 | 358.7+52.7 | 2.7 | 7.1 | |
| | | | 930719.2 | 341.9+58.3 | 7.5 | 5.1 | |
| | | | 930822.2 | 341.2+56.1 | 2.8 | 4.0 | |
| | | | 930831 | 351.3+47.5 | 1.7 | 7.0 | |
| 93. | 920325-<br>920718 | 351.8-30.6<br>(299.6-47.4) | 920325.1 | 349.6- 26.9 | 3.4 | 4.1 | $1 \cdot 10^{-6}$<br>4.4 |
| | | | 920406 | 339.2- 26.3 | 1.7 | 11.8 | |
| | | | 920428 | 5.5- 31.4 | 2.7 | 11.8 | |
| | | | 920525.2 | 357.9- 29.7 | 1.6 | 5.3 | |
| | | | 920605 | 356.3- 22.4 | 3.2 | 9.1 | |
| | | | 920613.1 | 341.8- 38.9 | 4.9 | 11.7 | |
| | | | 920618 | 354.1- 41.8 | 4.8 | 11.3 | |
| | | | 920708 | 349.3- 36.5 | 3.7 | 6.3 | |
| | | | 920718.2 | 341.7- 29.6 | 1.7 | 8.8 | |
| 94. | 930205-<br>930328 | 355.0-20.9<br>(286.9-42.2) | 930205 | 349.2- 15.9 | 6.2 | 7.4 | $4 \cdot 10^{-7}$<br>4.6 |
| | | | 930214.2 | 351.7- 18.6 | 2.6 | 3.8 | |
| | | | 930301 | 359.1- 26.9 | 1.9 | 7.0 | |
| | | | 930310 | 344.1- 22.0 | 1.8 | 10.2 | |
| | | | 930326.2 | 348.5- 12.8 | 3.9 | 10.2 | |
| | | | 930328 | 5.6- 24.1 | 2.8 | 10.3 | |